\documentclass[showpacs,amsmath,amssymb,twocolumn,superscriptaddress,notitlepage,preprintnumbers,pra,floatfix]{revtex4-1}

\usepackage[dvips]{graphicx}
\usepackage{amsmath,amssymb,amsthm,mathrsfs,amsfonts,dsfont}
 
\usepackage{dcolumn}
\usepackage{bm}
\usepackage{amsmath}
\usepackage{amssymb}
\usepackage{braket}
\usepackage{physics}
\usepackage{amsthm}
\usepackage{natbib}
\usepackage{mathtools}
\usepackage{diagbox}
\usepackage[utf8]{inputenc}
\usepackage[english]{babel}
\usepackage{scalerel}[2014/03/10]
\usepackage{stackengine}
\usepackage{algorithm}
\usepackage[noend]{algpseudocode} 
\usepackage{color}
\usepackage{orcidlink}

\usepackage{changes}

\usepackage{hyperref}
\hypersetup{colorlinks=true, linkcolor=blue, citecolor=blue, urlcolor=black }

\usepackage{qcircuit}

\graphicspath{{figures/}}

\newcommand{\ud}{\mathrm{d}}

\newcommand{\Id}{\ensuremath{\mathbf{I}}}

\newcommand{\III}{\mathcal{I}}
\newcommand{\SSS}{\mathcal{S}}

\begin{document}

\newcommand{\PKU}{Center on Frontiers of Computing Studies, Peking University, Beijing 100871, China}

\newcommand{\PKUCS}{School of Computer Science, Peking University, Beijing 100871, China}

\title{Simulating the Spread of Infection in Networks with Quantum Computers}

\author{Xiaoyang Wang\orcidlink{0000-0002-2667-1879} }
\affiliation{\PKU}
\affiliation{Center of High Energy Physics, Peking University, Beijing 100871, China}
\author{Yinchenguang Lyu}
\affiliation{Department of Modern Physics, University of Science and Technology of China, Hefei 230026, China}
\author{Changyu Yao}
\affiliation{\PKU}
\author{Xiao Yuan\orcidlink{0000-0003-0205-6545}}
\affiliation{\PKU}
\affiliation{\PKUCS}

\date{\today}

\begin{abstract}
We propose to use quantum computers to simulate infection spreading in networks. We first show the analogy between the infection distribution and spin-lattice configurations with Ising-type interactions. Then, since the spreading process can be modeled as a classical Markovian process, we show that the spreading process can be simulated using the evolution of a quantum thermal dynamic model with a parameterized Hamiltonian. In particular, we analytically and numerically analyze the evolution behavior of the Hamiltonian, and prove that the evolution simulates a classical Markovian process, which describes the well-known epidemiological stochastic susceptible and infectious (SI) model. A practical method to determine the parameters of the thermal dynamic Hamiltonian from epidemiological inputs is exhibited. As an example, we simulate the infection spreading process of the SARS-Cov-2 variant Omicron in a small-world network.
\end{abstract}

\maketitle

\section{Introduction}

The simulation of infection spreading in networks is a significant and widely studied problem. In epidemiology, modeling the spreading process in complex networks lies at the core of our understanding of infectious diseases~\cite{Satorras_15}. At the same time, the spreading of information, computer virus, and advertisement can be modeled as a contagion process. Thus, the effect of computer safeguarding~\cite{Wang_2000}, viral marketing~\cite{Jurvetson2000WhatEI} can be estimated by carrying out the similar spreading simulation. We focus on infection spreading simulation in this article, which aligns with the need for understanding the pandemic of SARS-Cov-2~\cite{WHO_reports}.

A continuous time infection spreading process in a network with constant transition rates can be described by the Markov chain theory~\cite{Castellano_10,Mieghem_09,Mieghem_13}. However, in general, the number of entries in the transition matrix of the Markov chain grows exponentially with the number of nodes in the network. Thus, the analyses are limited to very small graphs~\cite{Satorras_15}. For this reason, analyzing the evolution behavior of infection spreading in large networks are restricted to simplified approaches, such as the mean-field approach~\cite{Chakrabarti_2008}, where the fluctuation in the Markov chain is ignored, and Monte-Carlo methods~\cite{Lara2019AnalogyBT,Mello_2021,Crisostomo2012AnIM}, where the Markovian transition matrices simulate nearest-neighbour infection spreading processes. In this work, we propose to use quantum computers to simulate the Markov chain, such that more complicated infection spreading processes can  be  simulated including long-range and household transmission. We propose a thermal dynamic model with a parameterized Hamiltonian, such that the Markov chain can be simulated by the Schr\"odinger evolution of the Hamiltonian. Since the Schr\"odinger evolution can be carried out efficiently on quantum computers, this approach can be extended to large graphs.

Modeling infection spreading with quantum computers has been investigated in previous works~\cite{Britt_2020, Padhi_2020}. For example, in Ref.~\cite{Padhi_2020}, the authors propose to use the Heisenberg spin model to simulate infection spreading. However, the relationship between the evolution of the spin model and the infection spreading is not clear. Here, we propose a spin model simulating a thermal dynamic system. We provide both perturbative and numerical analyses on the evolution of our proposed spin model, and show that the evolution simulates the classical stochastic susceptible and infectious (SI) model~\cite{Lara2019AnalogyBT, Fred2008}.

As a demonstration, we implement our model to describe the spreading of SARS-Cov-2 variant Omicron~\cite{Omicron_transmissibility} in a small-world network. We provide a practical method to determine the free parameters of the proposed thermal dynamic Hamiltonian, where the secondary attack rate (SAR)~\cite{10.1001/jama.2022.3780} and basic reproduction number (BRN)~\cite{CHEN202069, Talha2022} of Omicron are utilized as epidemiological inputs. We compare the simulation results on quantum computers and the real-world experimental data, and show their qualitatively consistency.

This article is organized as follows. In section \ref{sec:phenomenology}, we review the phenomenological SI model and describe the physical infection spreading scenario that we attempt to simulate. In section \ref{sec:thermal-dynamic-model}, we map the infection spreading scenario to the thermal dynamic model, which is then shown to simulate the stochastic SI model using the perturbative analysis. In section \ref{sec:numerics}, by carrying out numerical simulations, we verify the perturbative analysis, show how to determine the free parameters of the thermal dynamic Hamiltonian, and compare the numerical results on quantum computers to data in the real world. Finally, we discuss possible extensions and other applications of our proposed method in section \ref{sec:discussion}.

\section{Phenomenology}\label{sec:phenomenology}
In epidemiology, the phenomenological susceptible and infectious (SI) model is usually used to predict the evolution of an epidemic~\cite{Padhi_2020, Mello_2021}. The human population in the model is divided into two compartments where $\SSS$ represents the susceptible population and $\III$ represents the infected population. These two quantities are related by two coupled first-order differential equations and can be solved given the initial conditions. One of the equations is
\begin{align}
    \frac{\ud \SSS}{\ud t}=-\beta\kappa \III \SSS,
    \label{eq:SI-interaction}
\end{align}
where $\beta$ is the contact transmission risk, and $\kappa$ is the average number of contacts between each individual in the two parts of the people. The contact transmission risk characterizes the virus transmissibility, and the average number of contacts reflects the contact pattern of human population, which is influenced by, for example, geographical distance and population distribution.

Description using Eq.~(\ref{eq:SI-interaction}) is coarse grained, because the average number of contacts $\kappa$ is an average of all the number of contacts between individuals. To describe the infection spreading in more detailed, we consider the simulation on a network. The network is consisted of nodes and edges. We assign each node $j\in \mathcal{N}$ to represent an individual or a group of people, such as a community or household members of the infected individual. The edges represent paths of infection spreading between the individuals or groups. Assume the total population on a node is $N_j,j\in \mathcal{N}$, which can be divided into susceptible population $S_j$ and infected population $I_j$ so that $N_j=S_j+I_j$. In our simulation, we assume the index patients are introduced simultaneously at a particular time, where we set $t=0$. To study the infection spreading from the index patients, we divide the nodes $\mathcal{N}$ into infectious nodes $I$ and susceptible nodes $S$. We assign one node to each index patient, so that the initial conditions of the evolution is $I_i(t=0)=N_i=1$ for the infectious nodes $i \in I$, and $S_j(t=0)=N_j$ for the susceptible nodes $ j \in S$.

We attempt to simulate the early stage of infection spreading in a network. In the early stage of the infection spreading, the exposed individuals have been infected by the disease but cannot yet transmit it. In other words, the secondary infection transmission will be ignored so that there is no edge between the susceptible nodes in the network. An illustration of the network is shown in figure~\ref{fig:virus-transmission}\textbf{a}, where the infection spreads from the infectious individual (red doll) to the susceptible group (blue dolls) along the spreading path (solid black line). 

Phenomenologically, infection spreading in such kind of networks can be described by differential equations
\begin{align}
    \frac{\ud S_{j}}{\ud t}=-\beta\sum_{i\in I}\kappa_{ij} I_i S_{j},\quad \forall  j\in S,
    \label{eq:si-differential-equations}
\end{align}
where $\beta$ is the contact transmission risk and $\kappa_{ij}$ is the averaged number of contacts in unit time between the infectious node $i$ and the susceptible node $j$. In the early stage of the spreading, the infectious node keep $I_i=1$, which is independent of time. Thus, Eq.~(\ref{eq:si-differential-equations}) can be solved explicitly 
\begin{align}
    P_{j} (t)\equiv \frac{S_{j} (t)}{S_{j} (0)} = \exp{-\beta\sum_{i\in I}\kappa_{ij}I_i t},\quad j\in S,
    \label{eq:phenomenology-single-exponential-decay}
\end{align}
where $P_{j}$ represents the \textit{survival probability} of the susceptible node $j$. 

Survival probability is the key quantity that we aim to estimate in this work. According to Eq.~(\ref{eq:phenomenology-single-exponential-decay}), the survival probability has an exponential decay behavior, where the decay rate is a function of $\beta$ and $\kappa_{ij}$. However, the phenomenological description is a mean-field approximation where the fluctuation in the stochastic process is ignored. The full description of the infection spreading process using the Markov chain will be discussed in the next section. Nonetheless, we will see that the exponential decay behavior indicated by Eq.~(\ref{eq:phenomenology-single-exponential-decay}) remains.

\section{Thermal dynamic model}\label{sec:thermal-dynamic-model}

In this section, we establish the thermal dynamic model to simulate infection spreading processes. The model is established in a step-by-step way. In the first subsection, we discuss the analogy between an individual, a classical spin and a quantum spin, i.e., qubit, by reviewing the analogy between an individual and a classical spin in literature, and critically discuss the analogy between a classical spin and a qubit. In the second subsection, we describe the analogy between individuals’ contacts in networks and couplings in Ising-type models. The Ising-type model is static, and the dynamics should be introduced. Thus, in the third subsection, we review the classical Markovian dynamics and establish the thermal dynamic model based on the Ising-type model. Finally, using a perturbative analysis, we show that the classical Markovian dynamics can be simulated by the thermal dynamic model.

\subsection{Analogy between individual and qubit}
Before making the analogy between an individual and a qubit, we first introduce the analogy between an individual and a classical spin. Some literature have made the analogy between an individual and a classical spin~\cite{Lara2019AnalogyBT, Mello_2021, Crisostomo2012AnIM}.
A classical spin $\sigma=1(-1)$ represents a susceptible(infectious) individual. Each spin is equipped with a probability distribution $\{p_0,p_1\}$ such that the spin $\sigma$ has probability $p_0$ to take the value $1$ and $p_1$ to take the value $-1$. $p_0,p_1$ satisfy the normalization condition $p_0+p_1=1$. So the individual has survival probability $\hat{P}=p_0$, which can be rewritten as 
\begin{align}
    \hat{P}=\frac{1+p_0-p_1}{2}.
    \label{eq:classical-survival-probability}
\end{align}
This form of survival probability has a natural correspondence by introducing qubits, as shown below.

The basic component of a quantum computer is qubit. One qubit can be in two states $\ket{0}$ and $\ket{1}$ like the states $0$ and $1$ for a classical bit. The difference between a qubit and a classical bit is that qubit can also be in a linear combination of the two states
\begin{align}
    \ket{\psi}\equiv\alpha \ket{0}+\beta\ket{1},
    \label{eq:linear-combination}
\end{align}
where $\alpha,\beta$ are complex numbers that satisfy the normalization condition $|\alpha|^2+|\beta|^2=1$. The special states $\ket{0}$ and $\ket{1}$ are known as \textit{computational basis states}, which form an orthonormal basis of the vector space. More information about qubit can be found in ~\cite{Nielsen2000}.

Qubit is very similar to the classical spin introduced above. In quantum computation, the state of classical spin $\sigma=1(-1)$ corresponds to $\ket{0}(\ket{1})$ when we measure an observable $Z\equiv\ket{0}\bra{0}-\ket{1}\bra{1}$. One will get two values for the corresponding two states
\begin{align}
   \bra{0}Z\ket{0} = 1,\quad \bra{1}Z\ket{1} = -1.
\end{align}
For the state in Eq.~(\ref{eq:linear-combination}), after observation using $Z$, it would be in state $\ket{0}$ with probability $|\alpha|^2$ and in state $\ket{1}$ with probability $|\beta|^2$. So the expectation value of $Z$ observable is $|\alpha|^2-|\beta|^2$. 

The probability distribution $\{|\alpha|^2, |\beta|^2\}$ is a purely quantum effect, which is distinguished from the classical probability distribution such as the distribution $\{p_0,p_1\}$ for a classical spin. Assume a classical distribution $\{p_i\}$ with normalization condition $\sum_i p_i=1$. To combine the purely quantum distribution with the classical distribution, Landau introduced the concept of density operator~\cite{Landau1981Quantum}. For one qubit case, its density operator can be generally written as
\begin{align}
    \rho=\sum_i p_i\ket{\psi_i}\bra{\psi_i}, 
\end{align}
where $\ket{\psi_i}=\alpha_i\ket{0}+\beta_i\ket{1}$ with $|\alpha_i|^2+|\beta_i|^2=1$. Notice that the pure quantum state in Eq.~(\ref{eq:linear-combination}) is a special case of the density operator if the probability distribution $\{p_i\}$ is like $\{p_1=1,p_2=0,p_3=0, \ldots\}$. The $Z$ expectation value of this density operator can be calculated according to 
\begin{equation}
\begin{aligned}
    \langle Z\rangle &= \Tr (\rho Z)\\
    &=\sum_i p_i(|\alpha_i|^2-|\beta_i|^2),
\end{aligned}
\end{equation}
which is a generalization to the expectation value $|\alpha|^2-|\beta|^2$. The probability of measuring state in $\ket{0}$ is 
\begin{align}
    \hat{P}= \frac{1+\langle Z\rangle}{2},
    \label{eq:survival-probability}
\end{align}
which is the quantum analogy of $p_0$ for the classical spin. We define $\hat{P}$ as the survival probability of the single qubit.

However, the difference between a qubit and a classical spin is that the classical spin can not be in a linear combination of states like in Eq.~(\ref{eq:linear-combination}). An individual also can not be in a state like $\ket{\mathrm{infected}}+\ket{\mathrm{susceptible}}$, yet allowed in quantum mechanics. This conflict will not appear if we choose a classical initial state and an appropriate evolution Hamiltonian. In the Supplementary Materials~\cite{supp}, we show that when a qubit is weakly coupled to a heat bath and when the initial state and measurement are computational basis state $\ket{0}/\ket{1}$, then during the Hamiltonian evolution, the qubit would behave like a classical spin. In other words, the qubit density operator during the evolution has the form
\begin{align}
\rho=p_0\ket{0}\bra{0}+p_1\ket{1}\bra{1},
\end{align}
where only classical probabilities are involved. Thus, the survival probability in Eq.~(\ref{eq:survival-probability}) reduces to the classical form of Eq.~(\ref{eq:classical-survival-probability}). This fact justify the reasonability of simulating classical infection spreading processes on quantum computers.

To summarize the above analogies, we list the conventions used in our calculation
\begin{equation}
\begin{aligned}
    \mathrm{infected}&\Leftrightarrow \ket{1}\Leftrightarrow Z=-1,\\
    \mathrm{susceptible}&\Leftrightarrow \ket{0}\Leftrightarrow Z=1.
\end{aligned}
\end{equation}

The above analogy between an individual and a qubit can be easily extended to multi-individual and multi-qubit cases. By assigning each node with a qubit, the survival probability of the individual on node $j$ can be calculated according to 
\begin{align}
    \hat{P_j}= \frac{1+\langle Z_j\rangle}{2}.
    \label{eq:survival-probability-on-j}
\end{align}

\subsection{Analogy between individuals' contacts and Ising coupling}

\begin{figure*}
    \centering
    \includegraphics[width = 0.90\textwidth]{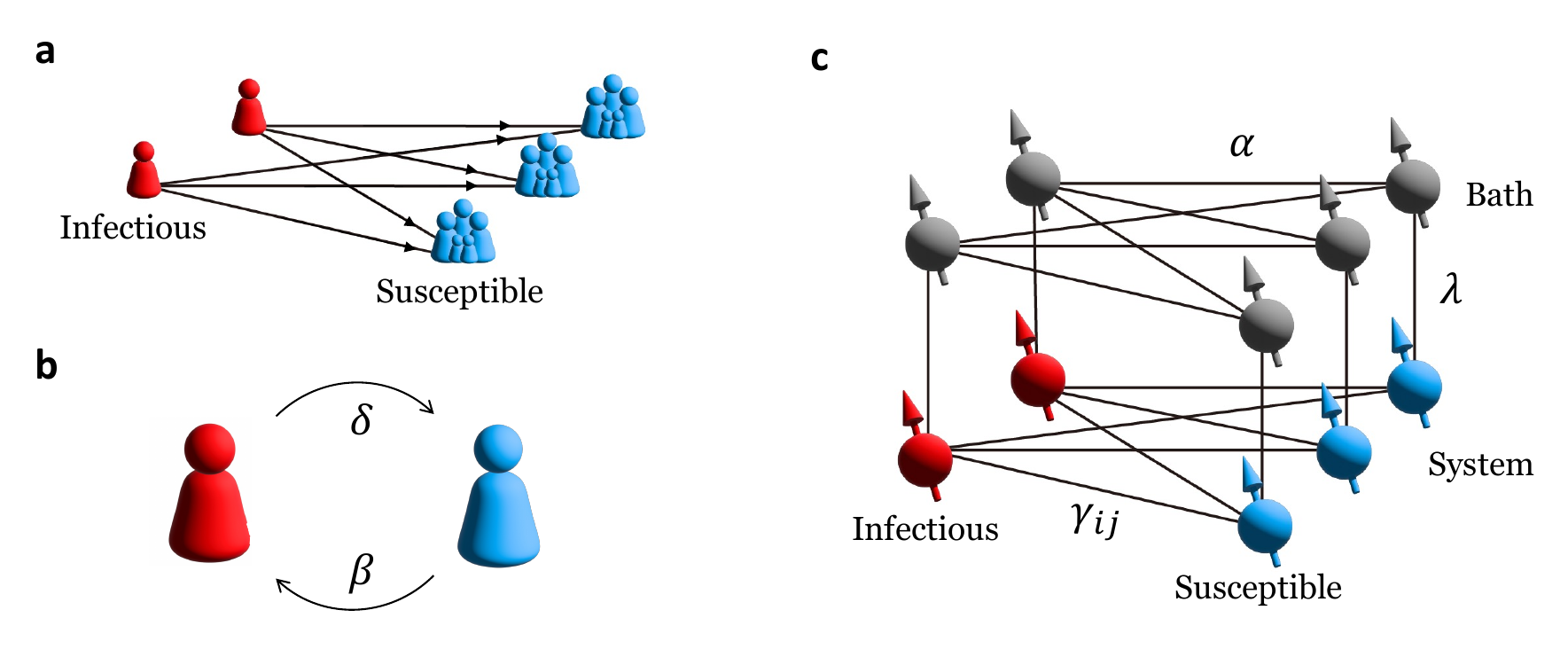}
    \caption{(\textbf{a}) An illustration of the scenario simulated in this work. The infection spread from the infectious individual (red doll) to the susceptible group (blue dolls) along the spreading path (solid black line). This network is translated to an Ising-type Hamiltonian, where we assign a spin for each node of the network and Pauli-$ZZ$ coupling for each edge.(\textbf{b}) An illustration of the Markovian process for one spin. We assume the susceptible individual has a constant probability $\beta$ getting infected and a constant probability $\delta$ recovered. (\textbf{c}) Visual demonstration of the thermal dynamic model. In the system layer, the red spins and blue spins correspond to the infectious individuals and susceptible groups in (\textbf{a}), respectively. Each spin in the system is accompanied by a grey spin in the bath layer. Horizontal lines represent Pauli-$ZZ$ coupling in the thermal dynamic Hamiltonian. Vertical lines represent Pauli-$XX$ coupling. The Latin letters $\gamma_{ij},\alpha,\lambda$ denote inter-system, inter-bath and system-bath coupling strength, respectively.}
    \label{fig:virus-transmission}
\end{figure*}

To describe individuals' contacts, we introduce Ising-type interaction to the system Hamiltonian
\begin{align}
    H_{s} = -\sum_{i\in I}\sum_{j\in S}\gamma_{ij}Z^s_i Z^s_j,
    \label{eq:system-hamiltonian}
\end{align}
where $\gamma_{ij}>0$ is the exchange interaction between nodes pair $(i,j)$. $\gamma_{ij}$ characterizes the contact frequency between the two individuals $i,j$, and we expect its correspondence to $\kappa_{ij}$ in Eq.~(\ref{eq:si-differential-equations}). In Eq.~(\ref{eq:system-hamiltonian}), the pairwise $ZZ$-coupling is assigned if there is an edge between the corresponding two nodes in the network. Thus, to simulate the infection spreading in networks such as the one shown in figure~\ref{fig:virus-transmission}\textbf{a}, we only consider $ZZ$-couplings between the infectious node $i\in I$ and susceptible node $j\in S$ in Eq.~(\ref{eq:system-hamiltonian}).

The Ising-type interaction used here is similar to the one introduced to the classical spin system~\cite{Lara2019AnalogyBT, Mello_2021, Crisostomo2012AnIM}. The difference is that, first, in the Hamiltonian Eq.~(\ref{eq:system-hamiltonian}), the infectious nodes connect to all the susceptible, which is distinguished from the nearest neighbor coupling in the references mentioned above. Thus, the long-distance infection spreading can be considered in our simulation. Secondly, since each infectious node has different coupling strength with the susceptible nodes, we can consider different geographical distances between the infectious and susceptible nodes. Moreover, complicated epidemic phenomena, such as household transmission and super-spreader, can also be investigated in this model. Thirdly, here we do not consider the secondary transmission. Thus, couplings among susceptible nodes $j\in S$ are ignored. However, secondary transmission is not fundamentally unachievable. The secondary transmission process can be simulated using a time-dependent Hamiltonian. For example, once a susceptible node becomes infectious during the evolution, i.e., the state $\ket{1}$ is measured, we can take this node from set $S$ to set $I$ so that the connection in the system Hamiltonian should be updated accordingly. In this work, the secondary transmission is omitted, and we only use the time-independent Hamiltonian.

The Ising-type Hamiltonian with $\gamma_{ij}>0$ corresponds to a ferromagnetic spin system. It has two degenerate ground states with all spins aligned to the same directions, i.e., all $\ket{0}$ or all $\ket{1}$. These two ground states correspond to the two possible final states after the infection spreading for an infinitely long time. If no index patient was introduced, the final state should be all $\ket{0}$. Otherwise, the final state should be all $\ket{1}$. Thus, one can expect to introduce a dynamics which is energy dissipative to transform an arbitrary initial state to the ground state. These considerations inspire us to introduce the following thermal dynamic model based on the Ising-type Hamiltonian.

\subsection{Thermal dynamics}

To make the spin system evolve, we should introduce dynamics to the system. On a classical computer, one can construct a Markovian process simulating the stochastic SI model~\cite{Lara2019AnalogyBT, Fred2008}. The so-called Markovian process indicates that the system's evolution is determined by its current state but not by its history. For example, the state of a single spin(an individual) is either susceptible or infected, which can be described by a binomial distribution $(p_0(t), p_1(t))^T$ at time $t$. We assume that the individual has a constant probability $\beta(\delta)$ getting infected(susceptible) when it is susceptible(infected), as depicted in figure~\ref{fig:virus-transmission}\textbf{b}. Thus, the dynamics of the single spin can be described by a Markovian transition matrix $S$
\begin{align}
    S=\left ( \begin{array}{cc}
        1-\beta & \delta  \\
        \beta & 1-\delta \end{array} \right),
        \label{eq:SI-model-markovian-matrix}
\end{align}
and the spin state in the following time time can be derived according to 
\begin{align}
    \left ( \begin{array}{c}
        p_{0}(t+1) \\
        p_{1}(t+1) \end{array} \right)=S\left ( \begin{array}{c}
        p_{0}(t) \\
        p_{1}(t) \end{array} \right).
        \label{eq:single-spin-markov-chain}
\end{align}
The transition matrix has two eigenvalues: $1$ and $1-\beta-\delta$, thus the survival probability of a single spin is 
\begin{align}
    p_0(t)=(1-\beta-\delta)^t (p_0(0)-p_0(\infty))+p_0(\infty),
    \label{eq:markov-chain-single-exponential-decay}
\end{align}
where $p_0(0)$ and $p_0(\infty)$ are the initial and final values of the survival probability. This expression indicates a single-exponential decay behavior when $\beta$ and $\delta$ are not very large, i.e., $\beta+\delta<1$, which is consistent with the mean-field results in the previous section.

In a network with $N$ nodes, the total number of states equals to $2^N$ for the two-compartment SI model, of which analysis is limited to very small graphs~\cite{Mieghem_09, Satorras_15} using classical computers. On the other hand, a quantum computer operates states living in Hilbert space, which has exponentially large dimensions with a growing number of qubits. Thus, quantum computer has the potential to simulate complicated infection spreading processes on large graphs. Here, we present a method simulating the Markovian processes using Schr\"odinger evolution on quantum computers.

On a quantum computer, the dynamics of a quantum system is described by the Schr\"odinger evolution of density operator
\begin{align}
    \rho_{sys}(t)=e^{-iH t}\rho_{sys}e^{iH t},
\end{align}
where $H=H_s$ is the Hamiltonian of the target system, and $\rho_{sys}=\rho_{sys}(0)$ is system's initial state. However, the Schr\"odinger evolution is energy conserving and reversible. Thus, we can not use Schr\"odinger evolution directly to simulate an energy-dissipative and irreversible spreading processes. 

One way to solve this problem is introducing additional qubits so that the target quantum system is a part of the whole system. We call the set of additional qubits as \textit{heat bath}, which is coupled with the target system. The Hamiltonian of the whole system can be generally written as
\begin{align}
    H = H_{s}\otimes \Id_{bath}+\Id_{sys}\otimes H_{b} +\lambda H_{sb},
\end{align}
where $H_{s}$ is the Hamiltonian of the target system, $H_{b}$ is the Hamiltonian of the heat bath, and $H_{sb}$ describes the interaction between the system and the bath. The interaction strength is characterized by a system-bath coupling strength $\lambda$. $\Id_{bath}$ and $\Id_{sys}$ are identity operators on the corresponding Hilbert space. With this extended Hamiltonian, the evolution of the system density operator is
\begin{align}
    \rho_{sys}(t) \equiv  \Tr_{b} (e^{-iHt}\rho_s\otimes\rho_{b,\beta} e^{iHt}),
    \label{eq:system-evolution-coupled-with-bath}
\end{align}
where $\rho_s,\rho_{b,\beta}$ are the initial state of the target system and the heat bath. $\Tr_{b}$ denotes partial trace over the bath Hilbert space.

The quantum evolution of the system according to Eq.~(\ref{eq:system-evolution-coupled-with-bath}) has been analyzed in Ref.~\cite{Terhal2000}. The authors perturbatively expanded the evolution Eq.~(\ref{eq:system-evolution-coupled-with-bath}) with small coupling strength $\lambda$. They have shown that, if we have an infinitely large heat bath under equilibrium with temperature $T$, the system will undergo a Markovian process up to the leading order of the perturbation, and finally, the system will evolve to the equilibrium state 
\begin{align}
    \rho_{sys}(t) \rightarrow \frac{e^{-H_{s}/ (k_B T)}}{\Tr (e^{-H_{s}/ (k_B T)})}
    \label{eq:thermal-state}
\end{align}
as $t$ is large enough. This equilibrium state has the same temperature as the heat bath, and $k_B$ is the Boltzmann constant. In the following content, we show that the Markovian process in the evolution can be used to simulate infection spreading, by firstly determining the heat bath temperature $T$ and give an explicit form of $H_b$ and $H_{sb}$. Then we show that the Markov chain Eq.~(\ref{eq:single-spin-markov-chain}) emerges from the determined Hamiltonian's evolution.

The bath temperature can be determined as follows. Notice that the temperature of the system's final state will be the same as the bath according to Eq.~(\ref{eq:thermal-state}). As mentioned in the previous subsection, we expect the system to evolve to its ground state, corresponding to  $T\rightarrow 0$ of the equilibrium state. Thus, in the evolution simulation, the bath temperature will initially be set to zero, i.e., set the bath spins to their ground state. The infinitely large heat bath can be simulated by resetting the bath spins to the ground state once after a specific time interval during the simulation~\cite{Terhal2000}.

It is left to choose an explicit form of the bath and interaction Hamiltonian. Theoretically, they could be arbitrary, to lead to the system's thermalization. So they are chosen mainly considering the simplicity of quantum simulation and analysis. The bath Hamiltonian is determined as follows. We expect that the thermalization for the system Hamiltonian is as fast as possible. It can be shown that the most rapid thermalization can be achieved if the energy differences in the system have the corresponding differences in the heat bath, as mentioned in \cite{Terhal2000}, also see the Supplementary Materials~\cite{supp}. It is physically reasonable. Because in that case, the energy emitted or absorbed by the system have the corresponding sink or source in the heat bath, so that the system resonate and interact with the bath intensively. Thus, a natural choice of the bath Hamiltonian follows the same form as the system Hamiltonian
\begin{align}
    H_{b} = -\sum_{i\in I}\sum_{j\in S} \alpha_{ij} Z^b_i Z^b_j,
    \label{eq:bath-hamiltonian}
\end{align}
where $Z^b_i$ is the Pauli-$Z$ operator on the bath Hilbert space and $\alpha_{ij}>0$ is a positive real number. With this form, the most rapid thermalization can be achieved by setting $\gamma_{ij}=\alpha_{ij}$. Further, we expect that the bath coupling reflects the contact transmission risk $\beta$ of a given kind of infectious disease, which should be uniform across all spins. Thus, all the bath couplings should be identical, which leads to
\begin{align}
    H_{b} = - \alpha\sum_{i\in I}\sum_{j\in S} Z^b_i Z^b_j.
    \label{eq:bath-hamiltonian-identical-bath}
\end{align}

Finally, we determine the form of the interaction Hamiltonian. We expect that the interaction Hamiltonian is local and uniform between the system and bath nodes. With locality, the Hamiltonian can be efficiently simulated on quantum devices. The uniformity requirement is because of the same reason as the one for the bath Hamiltonian. Thus, we choose the interaction Hamiltonian as
\begin{align}
    H_{sb} = - \sum_{i\in I\cup S} X^s_i X^b_i,
    \label{eq:system-bath-coupling-hamiltonian}
\end{align}
where $X^s_i, X^b_i$ are Pauli-$X$ operators acting on Hilbert space of the system node $i$ and the bath node $i$, respectively.

In summary, we propose a thermal dynamic Hamiltonian to simulate the infection spreading processes. The Hamiltonian reads
\begin{align}
    H =& -\sum_{i\in I}\sum_{j\in S} \gamma_{ij}Z^s_i Z^s_j\nonumber \\
        & -\lambda \sum_{i\in I\cup S} X^s_i X^b_i\nonumber \\
        & - \alpha \sum_{i\in I}\sum_{j\in S} Z^b_i Z^b_j,
        \label{eq:main-hamiltonian}
\end{align}
where $\gamma_{ij}, \lambda,\alpha$ are all positive real numbers. The superscripts $s,b$ denote the Hilbert space of the system and heat bath, respectively. A visual demonstration of spins and their interaction is shown in figure~\ref{fig:virus-transmission}\textbf{c}. The total number of qubits required to simulate this problem is $2\times  (|I|+|S|)$, where $|\cdot|$ denotes the number of elements in the set. All the spin interactions are local and thus can be simulated efficiently on near-term quantum computers. 

During the Schr\"odinger evolution, we also need \textit{reset} operation on the infectious and bath qubits. Resetting on the infectious qubits is to keep the patients in $\ket{1}$ state. Resetting the bath qubits is to simulate an infinitely large heat bath. More details about the reset operation on quantum computers can be found in the Supplementary Materials~\cite{supp}.

\subsection{Perturbative analysis}\label{sec:Perturbative analysis}
To see what Markovian process the spin system simulated using Hamiltonian Eq.~(\ref{eq:main-hamiltonian}), we carry out perturbative analysis using the technique proposed in Ref.~\cite{Terhal2000}. Here we present a simple network with one infectious node and one susceptible node. The general analysis is shown in the Supplementary Materials~\cite{supp}. The Hamiltonian for the simple network is written as
\begin{align}
    H =& -\gamma Z^s_0 Z^s_1\nonumber \\
        & -\lambda( X^s_0 X^b_0+ X^s_1 X^b_1)\nonumber \\
        & -\alpha Z^b_0 Z^b_1.
        \label{eq:simplified-two-nodes-main-hamiltonian}
\end{align}
Here, we allocate node 0 to the patient and node 1 to the susceptible. The dynamics is consisted of Schr\"odinger evolution $S_{\lambda,\Delta t}$ and resetting operation $\mathbf{R}$ on the patient qubit, where $\lambda$ is a small system-bath coupling and $\Delta t$ is the resetting time interval. So that the whole evolution is a staggering implementation $\mathbf{R}S_{\lambda,\Delta t}\mathbf{R}S_{\lambda,\Delta t}\ldots$. Both operations can be represented using a Markovian matrix on the susceptible node up to $\mathcal{O}(\lambda^4)$, and their multiplication is still a Markovian matrix
\begin{align}
    \mathbf{R}S_{\lambda,\Delta t}=\left ( \begin{array}{cc}
        1-\lambda^2 A_1 & \lambda^2 A_0  \\
        \lambda^2 A_1 & 1-\lambda^2 A_0 \end{array} \right)+\mathcal{O}(\lambda^4)
        \label{eq:single-s-markovian-matrix}
\end{align}
where 
\begin{align}
    A_0\equiv \frac{\sin^2 ( (\gamma+\alpha)\Delta t)}{ (\gamma+\alpha)^2},\quad A_1\equiv \frac{\sin^2 ( (\gamma-\alpha)\Delta t)}{ (\gamma-\alpha)^2}.
    \label{eq:A0-A1}
\end{align}
We see that both $A_0,A_1$ are functions of parameters in the Hamiltonian. The detailed derivation of these expressions can be found in the Supplementary Materials~\cite{supp}. 

Compare Markovian matrix Eq.~(\ref{eq:single-s-markovian-matrix}) with Eq.~(\ref{eq:SI-model-markovian-matrix}). We can identify that
\begin{align}
     \lambda^2 A_0\sim \delta, \quad  \lambda^2 A_1\sim \beta.
\end{align}
So that by carefully tuning the parameters in the Hamiltonian, the infection-recovery Markovian process with the infection (recovery) probability $\beta (\delta)$ can be simulated on quantum computers.

The perturbative analysis shows that our proposed thermal dynamic model simulates the stochastic SI model. A more complicated Markovian process involving immunity, secondary transmission and various node-node interaction can potentially be simulated using a more complicated time-dependent Hamiltonian. Those processes may not be easily simulated using Monte-Carlo algorithm~\cite{Lara2019AnalogyBT, Mello_2021, Crisostomo2012AnIM}. On the other hand, there are limitations of the method proposed here. For example, some epidemic models still cannot be simulated, such as the ones containing non-Markovian processes~\cite{Boguna_2014}.

Though the Markovian matrix shown in Eq.~(\ref{eq:single-s-markovian-matrix}) is derived perturbatively up to the leading order $\mathcal{O}(\lambda^2)$, in the numerical simulation, the value of $\lambda$ can be chosen arbitrarily so that the higher order correction can be neglected. It can be achieved by properly choosing the scale of the simulation time to the time in the real world. For example, in our following numerical simulations, we choose one computation time corresponding to one day in real life such that the simulated $\lambda$ is small enough and the perturbative results are consistent with the numerical simulation results. 


The inter-bath coupling $\alpha$ can be pre-determined using perturbative analysis results. We expect the infection rate to be zero as $\gamma=0$ in the system Hamiltonian. According to the expression $A_0, A_1$ in Eq.~(\ref{eq:A0-A1}), it can be achieved by setting $\alpha\Delta t=k\pi$ where $k$ is an arbitrary integer. In the numerical simulations, we choose $\Delta t=1$ and $k=1$ so that $\alpha=\pi$.

\section{Numerical simulation and parameters determination}\label{sec:numerics}

\begin{figure}
    \centering
    \includegraphics[width = 0.45\textwidth]{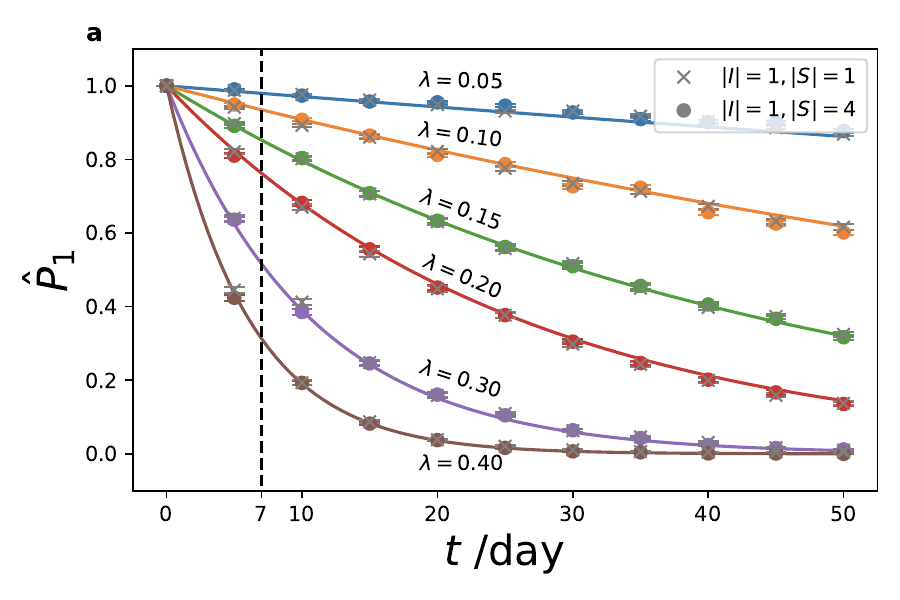}
    \includegraphics[width = 0.45\textwidth]{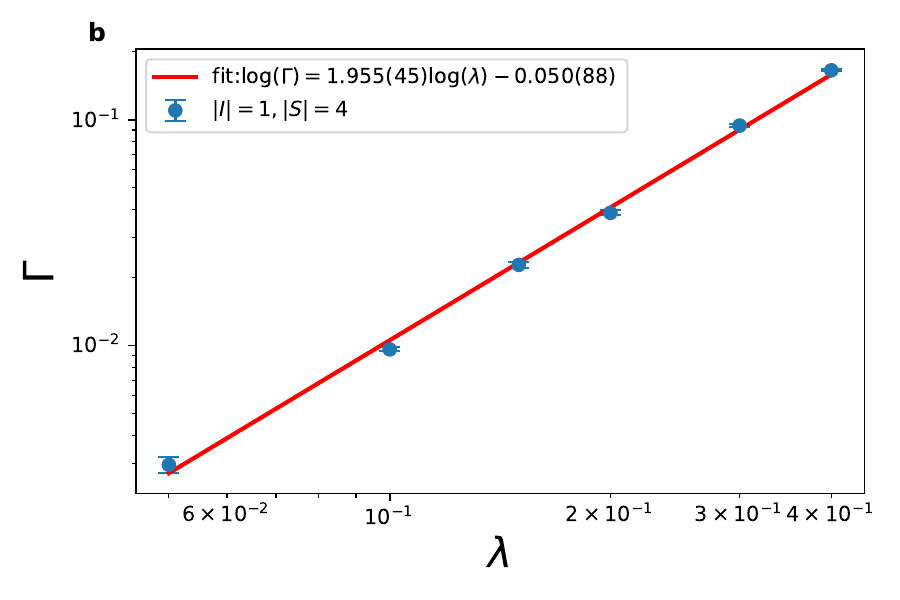}
    \caption{Single-exponential decay behavior of the survival probability and the determination of system-bath coupling $\lambda$.  (\textbf{a})Numerical results for survival probabilities of the household node $j=1$ as a function of time, where the time is in the day unit, under different system-bath coupling $\lambda$. The curves exhibit single-exponential decay $\exp (-\Gamma t)$, where $\Gamma$ is the infection rate. The simulation is carried out under two lattice size: $|I|=1,|S|=4$ and $|I|=1,|S|=1$. The vertical black-dashed line labels the seventh day.  (\textbf{b})The household node's infection rate $\Gamma$ as a function of coupling $\lambda$, under log-log plot. $\Gamma$ can be well fitted by a straight line with the slop $1.955 (45)\simeq 2$. It indicates the quadratic behavior $\Gamma\simeq \lambda^2$. With the input $\mathrm{SAR}_{\mathrm{Omicron}}=25.1\%$, we find $\lambda=0.201 (16)$. The number in the parentheses is the fitting error of the last two displayed digits.}
    \label{fig:infection-rate-system-bath-coupling}
\end{figure}

\begin{figure}
    \centering
    \includegraphics[width = 0.45\textwidth]{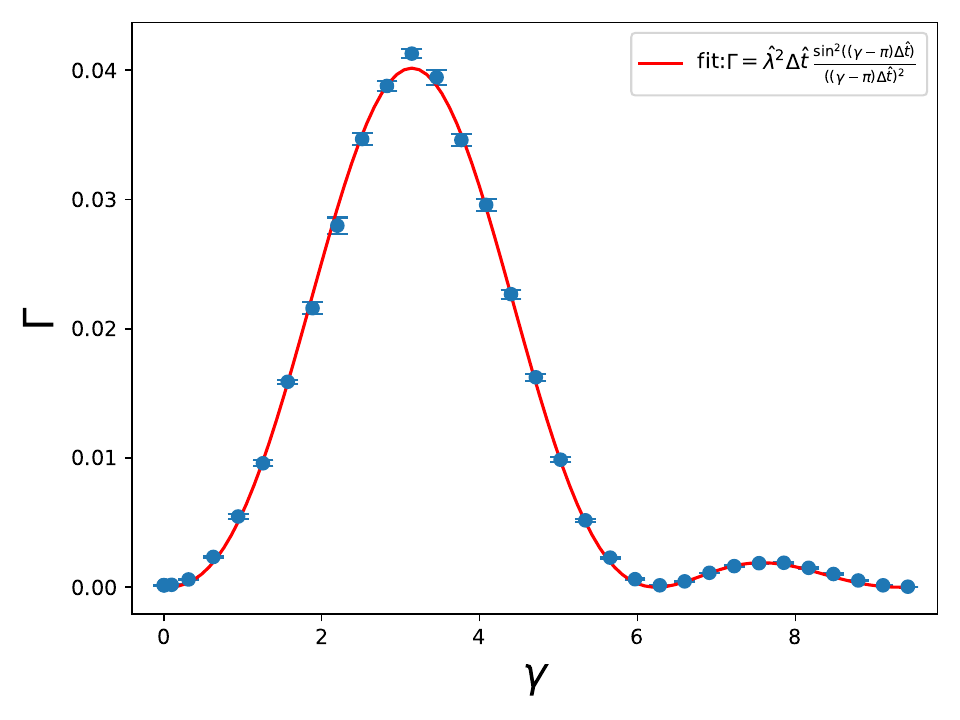}
    \caption{The infection rate $\Gamma$ as a function of inter-system coupling $\gamma$. Each data point is derived by fitting a single-exponential decay curve as in figure~\ref{fig:infection-rate-system-bath-coupling}. The error bar of each point is the fitting error. The curve can be well fitted by a sinc-function $\Gamma (\gamma)=\hat{\lambda}^2 \Delta  \hat{t} \sin^2 ( (\gamma-\pi) \Delta \hat{t})/ ( (\gamma-\pi) \Delta  \hat{t})^2$ with undetermined parameters $\hat{\lambda}$ and $\Delta \hat{t}$. The fitting results are $\hat{\lambda}=0.1990 (5)$ and $\Delta \hat{t}=1.014 (5)$.}
    \label{fig:infection-rate-system-coupling}
\end{figure}

We carry out numerical simulations using the thermal dynamic model introduced in the previous section. The Schr\"odinger evolution is simulated on CNOT-based quantum circuits using the first-order Trotterization algorithm~\cite{Nielsen2000}, and the quantum circuits are executed on the quantum simulator using the Qiskit SDK~\cite{Qiskit}. Details on the Hamiltonian simulation can be found in the Supplementary Materials~\cite{supp}. 

In this section, we first check the perturbative results derived in the previous section. Then we show the procedure of free parameters determination for the thermal dynamic model, which simulates a given kind of infection spreading in a given community environment. As a demonstration, we simulate the spreading of disease caused by the SARS-Cov-2 variant Omicron~\cite{Omicron_transmissibility}. Other kinds of infectious diseases can be simulated in the similar way. We use Omicron's secondary attack rate (SAR) and basic reproduction number (BRN) as experimental inputs to determine the free parameters in the Hamiltonian. In the end of this section, we compare the simulation results of the determined Hamiltonian with the real-world experimental data. 

\subsection{Check perturbative results}

We firstly check the single-exponential decay behavior as predicted by the Markovian processes in section~\ref{sec:Perturbative analysis}. We carry out numerical simulation using Hamiltonian as shown in Eq.~(\ref{eq:simplified-two-nodes-main-hamiltonian}), with $\alpha=\pi$ and resetting time interval $\Delta t=1$. In Figure~\ref{fig:infection-rate-system-bath-coupling}\textbf{a} with $|I|=1,|S|=1$, we plot the estimated survival probabilities $\hat{P_j} (t)$ of the susceptible node $j=1$ for various system-bath coupling $\lambda$. The quantum evolution is carried out using first-order Trotter decomposition with Trotter step length $\delta t=0.01$. The expectation value is measured by running the quantum circuit 4096 times for each data point. The error bar is the statistical error (We use the same setting in the following simulations if not specified). The survival probabilities have single-exponential decay behavior depending on time $t$. This behavior is consistent with the mean-field result in Eq.~(\ref{eq:phenomenology-single-exponential-decay}).

To check the parameter dependence of the Markovian matrix in Eq.~(\ref{eq:single-s-markovian-matrix}), we define the \textit{infection rate} $\Gamma_j$ for node $j$ that is given by
\begin{align}
     \Gamma_j = -\frac{\partial \hat{P_j} (t)}{\partial t}\Big|_{t=0}.
     \label{eq:single-exp-decay}
\end{align}
In case the survival probability has single-exponential decay behavior, we have $\hat{P_j} (t)=e^{-\Gamma_j t}$. According to the Markovian matrix Eq.~(\ref{eq:single-s-markovian-matrix}), the infection rate of node $j=1$ should be
\begin{align}
    \Gamma (\gamma)\simeq \lambda^2\Delta t \left (\frac{\sin ( (\gamma-\pi)\Delta t)}{ (\gamma-\pi)\Delta t}\right)^2.
    \label{eq:infection-rate-analytic}
\end{align}
The notation $\simeq$ denotes that high order perturbation and the recovery probability $A_0$ are ignored. Because $A_0$ is very small compared to $A_1$ under the current parameters settings. Figure~ \ref{fig:infection-rate-system-coupling} plots the infection rate as a function of inter-system coupling $\gamma$. Each data point is derived by fitting a single-exponential decay curve, and the infection rate can be extracted. The data points are fitted using the analytic formula Eq.~(\ref{eq:infection-rate-analytic}). In this simulation, we take system-bath coupling $\lambda=0.201$. We see that $\Gamma$ is peaked at $\gamma=\pi$ and the curve is well fitted by a sinc-function with fitted parameters $\hat{\lambda}=0.1990 (5)$ and $\Delta \hat{t}=1.014 (5)$. They are consistent with theoretical values $\lambda=0.201$ and $\Delta t=1$. The little inconsistency at the peak $\gamma=\pi$ is mainly because the recovery probability $A_0$ is ignored in the fitting formula.

\subsection{Determination of free parameters in the Hamiltonian}\label{s-b-coupling}

This subsection demonstrates how to determine the free parameters in the thermal dynamic Hamiltonian for a specific infectious diseases in a small-world network. Then we can use the determined parameters to simulate more complicated networks.

\subsubsection{Specificity of Omicron}
 The infection spreading we attempt to simulate is the disease caused by Omicron. Omicron is a variant of SARS-CoV-2 first reported in South Africa on 24th November 2021. It has higher transmissibility and a lower case fatality rate compared to the earlier mutants of SARS-CoV-2. We use Omicron's secondary attack rate (SAR) and basic reproduction number (BRN) as epidemiological inputs to determine the free parameters in the Hamiltonian.

SAR is defined as the number of non-index susceptible household members with a positive test result within seven days after the sample date of the index case, divided by the total number of non-index household members~\cite{10.1001/jama.2022.3780}. Thus, it is the household infection probability $1-\hat{P}_j~(t=7)$. In Ref.~\cite{10.1001/jama.2022.3780}, the authors reported SAR for Omicron variants of SARS-Cov-2 in Norwegian households, where $\mathrm{SAR}_{\mathrm{Omicron}}=25.1\%$. We use this input to determine the system-bath coupling in the thermal dynamic Hamiltonian.

BRN is defined as the average number of susceptible populations generated by one contagious person~\cite{CHEN202069}. BRN is usually denoted by $R_0$. It is an essential epidemiological indicator to characterize the virus transmissibility. If $R_0<1$, the virus will decline and eventually disappear. If $R_0=1$, it will stay alive but will not be an epidemic. If $R_0>1$, it will cause an epidemic or even a pandemic. In \cite{Ying_2022}, the author collected the data from 1st November 2021 to 9th February 2022 and estimated that the average BRN of Omicron is around 9.5. Additionally, we assume the index patient is quarantined and gets non-contagious as soon as the patient experience symptoms. Thus, we treat $R_0$ as the total number of infected population by one index patient during the incubation period. The incubation period of Omicron is around four days, according to \cite{Jansen2021}.

\subsubsection{A small-world network and contact pattern}

\begin{figure}
    \centering
    \includegraphics[width = 0.45\textwidth]{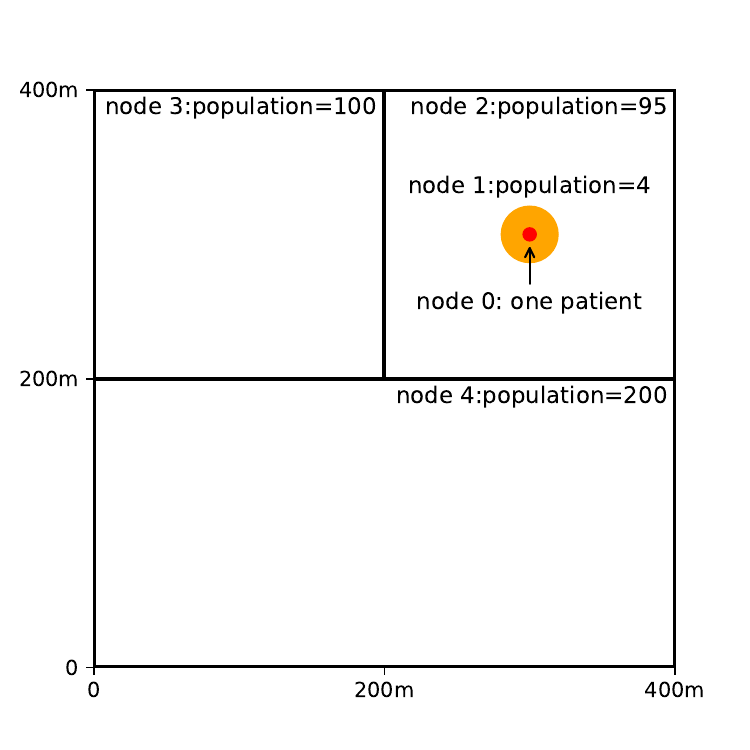}
    \caption{The map and population of a community environment. The community environment has five nodes: one for the index patient  (red dot, node 0), one for the patient's household members  (orange plate, node 1), and three for communities  (rectangles, nodes 2,3,4). The population in each community is assumed to be uniformly distributed.}
    \label{fig:unit-locations}
\end{figure}

We assume a toy model that contains a small-world network simulating a community environment and a particular contact pattern, such that the SAR and BRN describe the infection spreading within this community environment. The small-world network consists of five nodes: one for the index patient (at node 0), one for the patient's household members (at node 1), and three for communities (at nodes $2,3,4$). So that we have $|I|=1,|S|=4$, and their geographical distribution is shown in figure~\ref{fig:unit-locations}. The corresponding thermal dynamic Hamiltonian reads
\begin{align}
    H =& -\sum_{j=1}^4 \gamma_{0j}Z^s_0 Z^s_j\nonumber \\
        & -\lambda \sum_{j=1}^4 X^s_j X^b_j\nonumber \\
        & -\pi \sum_{j=1}^4 Z^b_0 Z^b_j.
        \label{eq:simplified-main-hamiltonian}
\end{align}
where $i=0$ is the infectious node and $j=1,2,3,4$ are susceptible.

The contact pattern describes the average number of contacts $\kappa_{ij}$ in the community environment, which is determined by factors such as the population distribution, geographical distance and environmental conditions, like temperature and humidity. In the toy model, we consider geographical distance of the community environment provided above, as well as the population distribution and the patient's daily track. We assume the population in each community is evenly distributed, and the daily track of the index patient follows an isotropic 2-d Gaussian distribution, which is centered at the patient's house. Thus, the average number of contacts in the community environment can be parametrized as
\begin{align}
    \kappa_{0j}=\kappa_{0}
    \frac{\int\!\!\!\int_{S_j}\ud^2 r\exp (-\frac{|\vec{r}-\vec{r}_0|^2}{2\sigma^2})}{\int\!\!\!\int_{S_j}\ud^2 r}, j=1,2,...|S|
    \label{eq:infection-rate-distance}
\end{align}
where $\vec{r}_0$ is the location of the patient's house, and $\sigma$ can be regarded as a distance scale of the patient's range of activity. The double integration is on the rectangle's area $S_j$ for node $j$, as shown explicitly in figure~\ref{fig:unit-locations}. As the area of the household node $S_1$ is infinitely small, we define $\kappa_{01}=\kappa_{0}$.

With the contact pattern described above, we can parameterize the inter-system coupling $\gamma_{0j}$ in Hamiltonian Eq.~(\ref{eq:simplified-main-hamiltonian}) with the parameter $\sigma$. According to the phenomenology, the infection rate and the average number of contacts are related by 
\begin{align}
    \Gamma_j=\beta\kappa_{0j}.
\end{align}
On the other hand, according to the perturbative results, the infection rate and inter-system coupling have the relationship
\begin{align}
    \Gamma_j\simeq \lambda^2 \left (\frac{\sin(\gamma_{0j}-\pi)}{ \gamma_{0j}-\pi}\right)^2.
    \label{eq:perturbative-infection-rate}
\end{align}
where we have chosen $\Delta t=1$. Combine the above two formulae, one finds
\begin{align}
     \lambda^2 \left (\frac{\sin (\gamma_{0j}-\pi)}{\gamma_{0j}-\pi}\right)^2\simeq \beta \kappa_{01} \frac{\int\!\!\!\int_{S_j}\ud^2 r\exp (-\frac{|\vec{r}-\vec{r}_0|^2}{2\sigma^2})}{\int\!\!\!\int_{S_j}\ud^2 r}.
     \label{eq:gamma-phenomenology-interpretation}
\end{align}
Thus, we can find an one-to-one correspondence of $\lambda$ and $\gamma_{0j}$ in the Hamiltonian to $\beta \kappa_{01}$ and $\sigma$. However, notice that the sinc-function is not monotonic for $\gamma_{0j}>0$. In our simulation, we choose $\gamma_{0j}$ within the monotonic domain $ (0,\pi]$, and we use the maximum point $\gamma_{01}=\pi$ to describe the interaction between the index patient with the household members. Thus, the system-bath coupling $\lambda$ depend on $\beta\kappa_{01}$ as
\begin{align}
    \lambda^2\simeq \beta\kappa_{01}.
    \label{eq:lambda-phenomenology-interpretation}
\end{align}
We see that $\lambda$ explicitly describes the transmissibility of the infectious diseases, which is consistent with our expectation when we heuristically determine the explicit form of the thermal dynamic Hamiltonian. Taking Eq.~(\ref{eq:lambda-phenomenology-interpretation}) into Eq.~(\ref{eq:gamma-phenomenology-interpretation}), the inter-system coupling can be parameterized as a function of $\sigma$
\begin{align}
    \left (\frac{\sin (\gamma_{0j}-\pi)}{\gamma_{0j}-\pi}\right)^2\simeq \frac{\int\!\!\!\int_{S_j}\ud^2 r\exp (-\frac{|\vec{r}-\vec{r}_0|^2}{2\sigma^2})}{\int\!\!\!\int_{S_j}\ud^2 r}.
     \label{eq:inter-system-coupling-sigma}
\end{align}
Thus, the inter-system couplings $\gamma_{0j}\in(0,\pi]$ are parameterized by one variable $\sigma$. 

$\gamma_{0j}$ also has a phenomenological interpretation: the deviation $(\pi-\gamma_{0j})$ reflects the average geographical distance of the community to the index patient. When $\gamma_{0j}$ goes to $0$, the distance goes to infinitely large. Thus, the long-distance transmission can be simulated using the thermal dynamic model.

\subsubsection{Determination of free parameters}

\begin{figure}
    \centering
    \includegraphics[width = 0.45\textwidth]{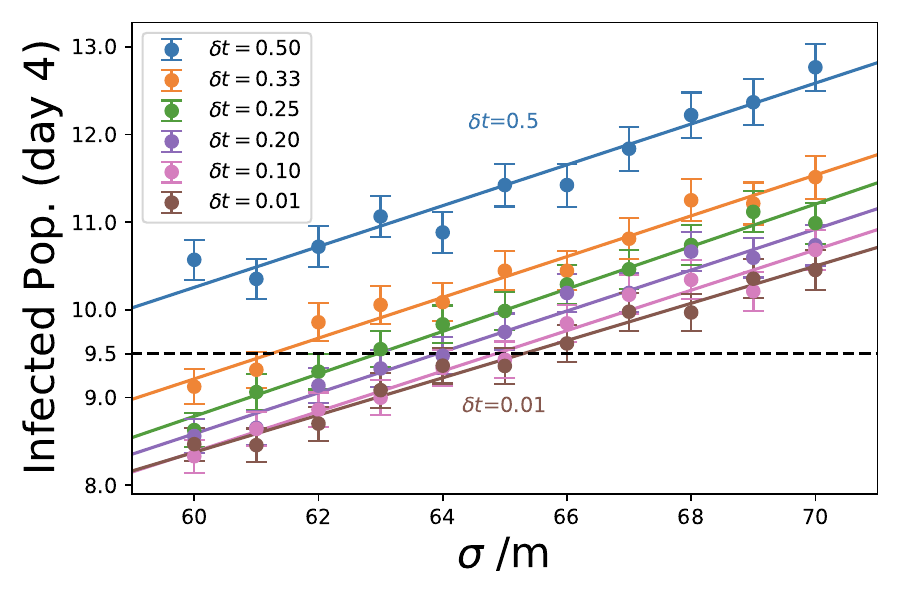}
    \caption{The total infected population 4 days after introducing one index patient into the community environment, as a function of distance scale $\sigma$. We use the first-order Trotter decomposition with decreasing Trotter step $\delta t$ until the results converge to a decent precision. We use a linear function to fit the curve. The black dashed line denotes the $R_0$ of Omicron. With $\delta t=0.01$, we find the required distance scale $\sigma=65 (3)$.}
    \label{fig:5-5-simulation-for-r0}
\end{figure}

\begin{figure}
    \centering
    \includegraphics[width = 0.45\textwidth]{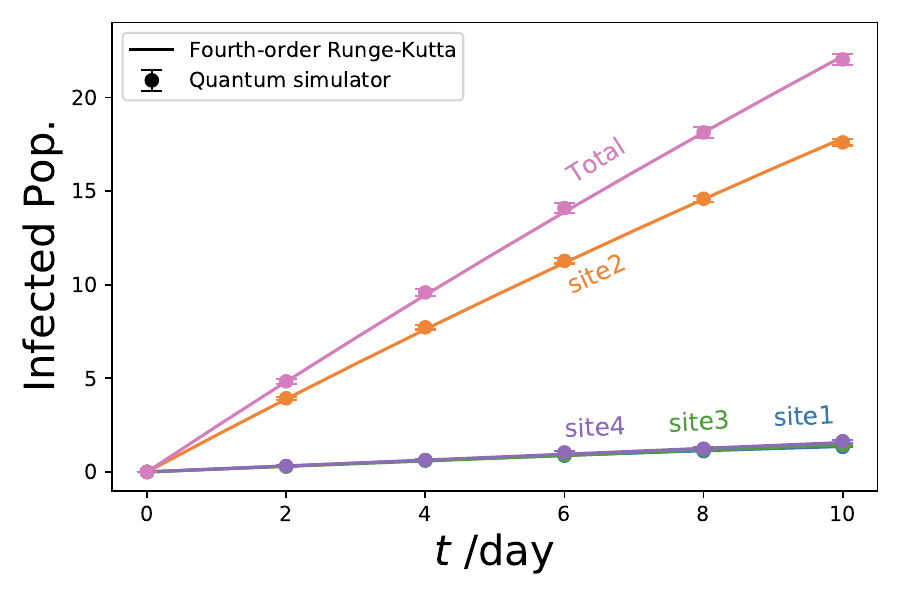}
    \caption{The infected populations of each susceptible node on different days of the community environment in figure~\ref{fig:unit-locations}. Here we use the distance scale $\sigma=65$. The total infected population by the introduced index patient on day 4 is $9.58 (20)$, which is consistent with the requirement of the $R_0$ of Omicron. We also plot the infected populations evaluated utilizing the fourth-order Runge-Kutta method. Results from the two methods converge within the error of statistics}
    \label{fig:5-5-simulation-check}
\end{figure}

\begin{figure*}[t]
  
  \includegraphics[width=2.0\columnwidth]{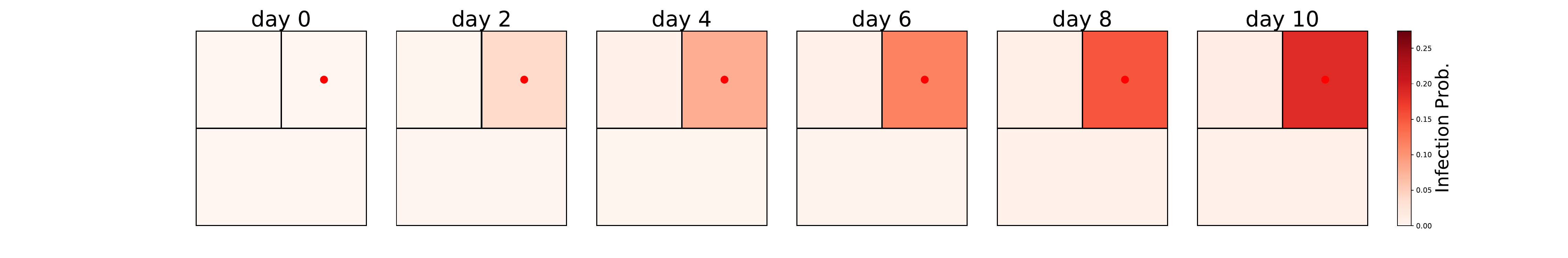}
  \includegraphics[width=2.0\columnwidth]{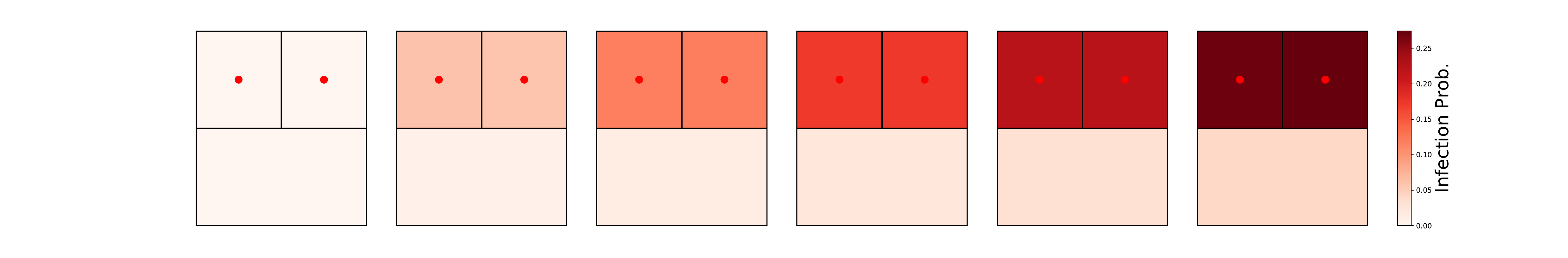}
  \caption{ Simulation of infection probabilities on different days, with one (upper row) and two (lower row) index patients introduced. The red points label the position of the index patients. The rectangles denote communities. We assume all the patients' activity ranges follow the same Gaussian distribution with $\sigma=65$. Each community's population and geometry settings follow figure~\ref{fig:unit-locations}. We see that the infection rates rise by introducing one more patient.}
  \label{fig:one-two-index-patients} 
\end{figure*}

With the given community environment and contact pattern, the Hamiltonian Eq.~(\ref{eq:simplified-main-hamiltonian}) has two free parameters $\lambda$ and $\sigma$. In this subsection, we determine these two free parameters with the SAR and BRN of Omicron.

SAR can determine the system-bath coupling $\lambda$ because both quantities are related to the survival probability of the household node. We can calculate the survival probability of the household node using the evolution of Hamiltonian  Eq.~(\ref{eq:simplified-main-hamiltonian}). Then by tuning $\lambda$, we can find a suitable value that satisfies the requirement of the SAR of Omicron. The simulation results are shown in figure~\ref{fig:infection-rate-system-bath-coupling} with $|I|=1,|S|=4$. In the upper panel, the survival probabilities explicitly show the single-exponential decay behavior for various values of $\lambda$. Recall that SAR is the household infection probability in the seventh day, which is denoted by the vertical black-dashed line in the figure. We plot the infection rate in the lower panel as a function of $\lambda$ shown in figure~\ref{fig:infection-rate-system-bath-coupling}\textbf{b}. It is a log-log plot and the slop $1.955(45)\simeq 2$ is consistent with the quadratic behaviour as predicted by Eq.~(\ref{eq:perturbative-infection-rate}) 

The epidemiological input $\mathrm{SAR}_{\mathrm{Omicron}}=25.1\%$ indicates the infection rate of the household member
\begin{align}
    \Gamma_1= -\frac{1}{t}\ln (1-\mathrm{SAR}_{\mathrm{Omicron}})\Big|_{t=7}.
\end{align}
Thus, we can solve for $\lambda$ utilizing the fitting formula in figure~\ref{fig:infection-rate-system-bath-coupling}\textbf{b} by requiring $\Gamma=\Gamma_1$. 
Thus, we have the system-bath coupling $\lambda=0.201 (16)$. This value has been adopted when we check the perturbative results in the previous subsection. To guarantee that the Trotter error and finite measurement error are not significant in the simulation, the same simulation results in figure~\ref{fig:infection-rate-system-bath-coupling} are checked using the fourth-order Runge-Kutta method as shown in the Supplementary Materials~\cite{supp}.

The distance scale $\sigma$ can be determined using $R_0$ of Omicron. Given a community environment and the population at each community $N_j$ as shown in figure~\ref{fig:unit-locations}, $R_0$ equals the total number of infected population by one patient after the incubation period (4 days). For different values of $\sigma$, the total number of the infected population is different and can be calculated using quantum computers. Then we match the results with $R_0$ of Omicron. The simulation results are shown in figure~\ref{fig:5-5-simulation-for-r0}. To control the systematic error from Trotter decomposition, we decrease Trotter steps until the results converge within the error of statistics. The total infected population is derived from survival probabilities
\begin{align}
    \mathrm{Infected~Pop.}=\sum_{j=1}^4 N_j(1-\hat{P}_j),
\end{align}
where the survival probabilities are measured by running the quantum circuit 50000 times for each data point, and the error bar is the statistical error. We see that the measured infected populations converge as $\delta t=0.01$. Taking this Trotter step, we fit the data by a linear function and find the distance scale $\sigma=65 (3)m$ by requiring $R_0=9.5$. With this distance scale, in more complicated situations, all the inter-system couplings of the thermal dynamic Hamiltonian can be determined according to Eq.~(\ref{eq:infection-rate-distance}).

In figure~\ref{fig:5-5-simulation-check}, we check the total number of infected populations with the Hamiltonian determined above. We find the total number of the infected population is $9.58 (20)$ on the fourth day, which is consistent with the requirement of $R_0$ of Omicron. We also use the fourth-order Runge-Kutta method to carry out the Hamiltonian simulation, as shown in the figure. The fourth-order Runge-Kutta method has less systematic error than the quantum simulator but can only simulate small networks. We see that the results of the infected population from quantum simulators are consistent with those of the fourth-order Runge-Kutta method within the error of statistics. More details of the Runge-Kutta method are shown in the Supplementary Materials~\cite{supp}.

\subsubsection{Simulation for two infectious nodes}

We can simulate the thermal dynamic model having more than one infectious node, using the system-bath coupling $\lambda$ and distance scale $\sigma$ as determined above. Apart from the difference in the $ZZ$-coupling of the Hamiltonian, for more than one infectious node, the inter-bath coupling $\alpha$ should differ from the model having one infectious node. According to the perturbative analysis (See details in the Supplementary Materials~\cite{supp}), the bath coupling should be $\alpha=\pi/|I|$. In figure~\ref{fig:one-two-index-patients}, we plot the infection probabilities of each community on different days, introducing one and two index patients, respectively. We see that the infection probabilities in the case of introducing two patients are larger than that of introducing one. The inter-bath coupling is $\alpha=\pi$ for the one patient case and $\alpha=\pi/2$ for the two patients case.

\subsection{Comparison of the numerical results to data in the real world}

\begin{figure}
\centering
  \includegraphics[width = 0.45\textwidth]{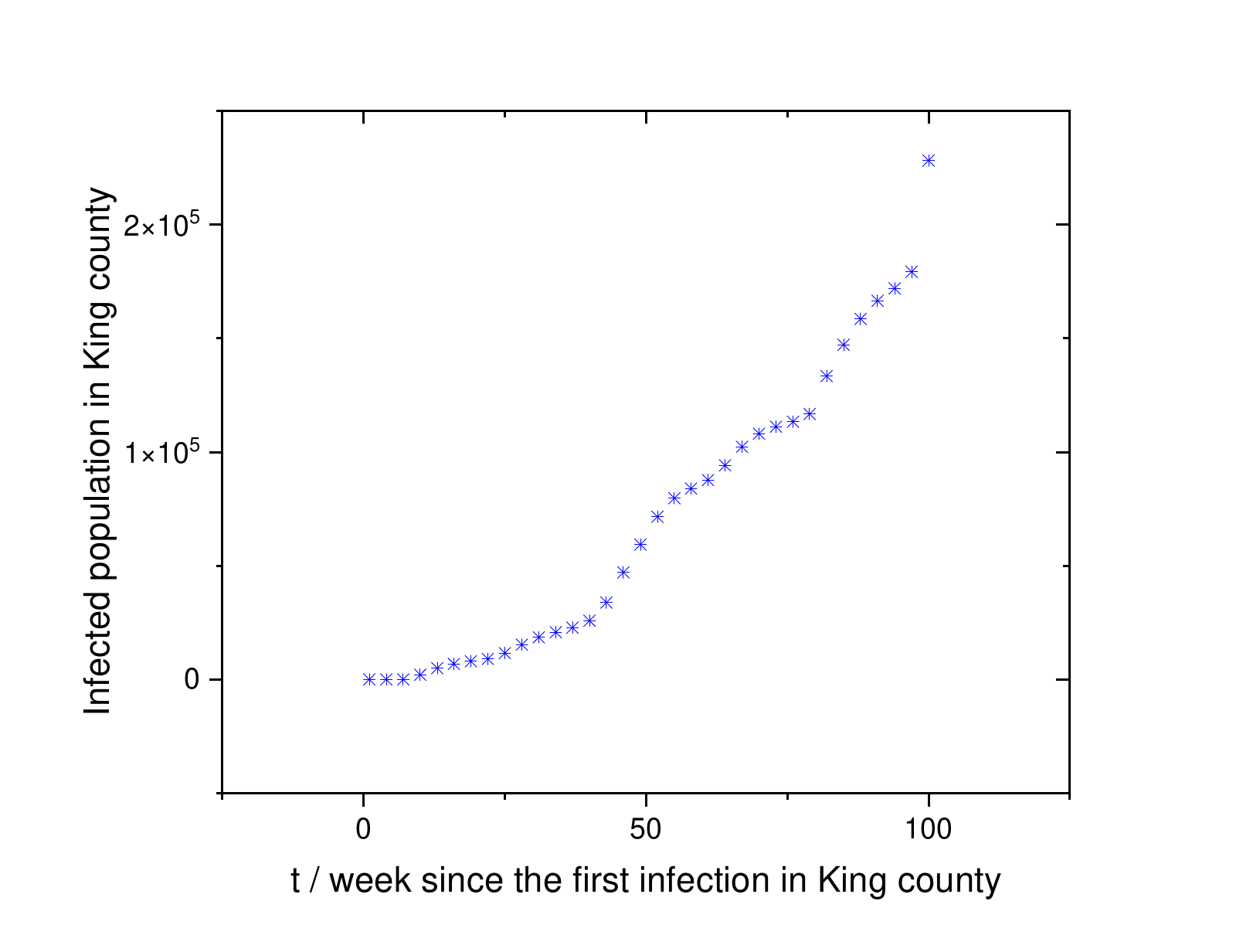}
  \caption{Infected population by weeks since the first infected case in King county, a state of Washington. Data recorded is from 22nd January 2020 to 27th February 2022.}
  \label{fig:infected-population-King} 
\end{figure}

\begin{figure}[b]
\centering
  \includegraphics[width = 0.45\textwidth]{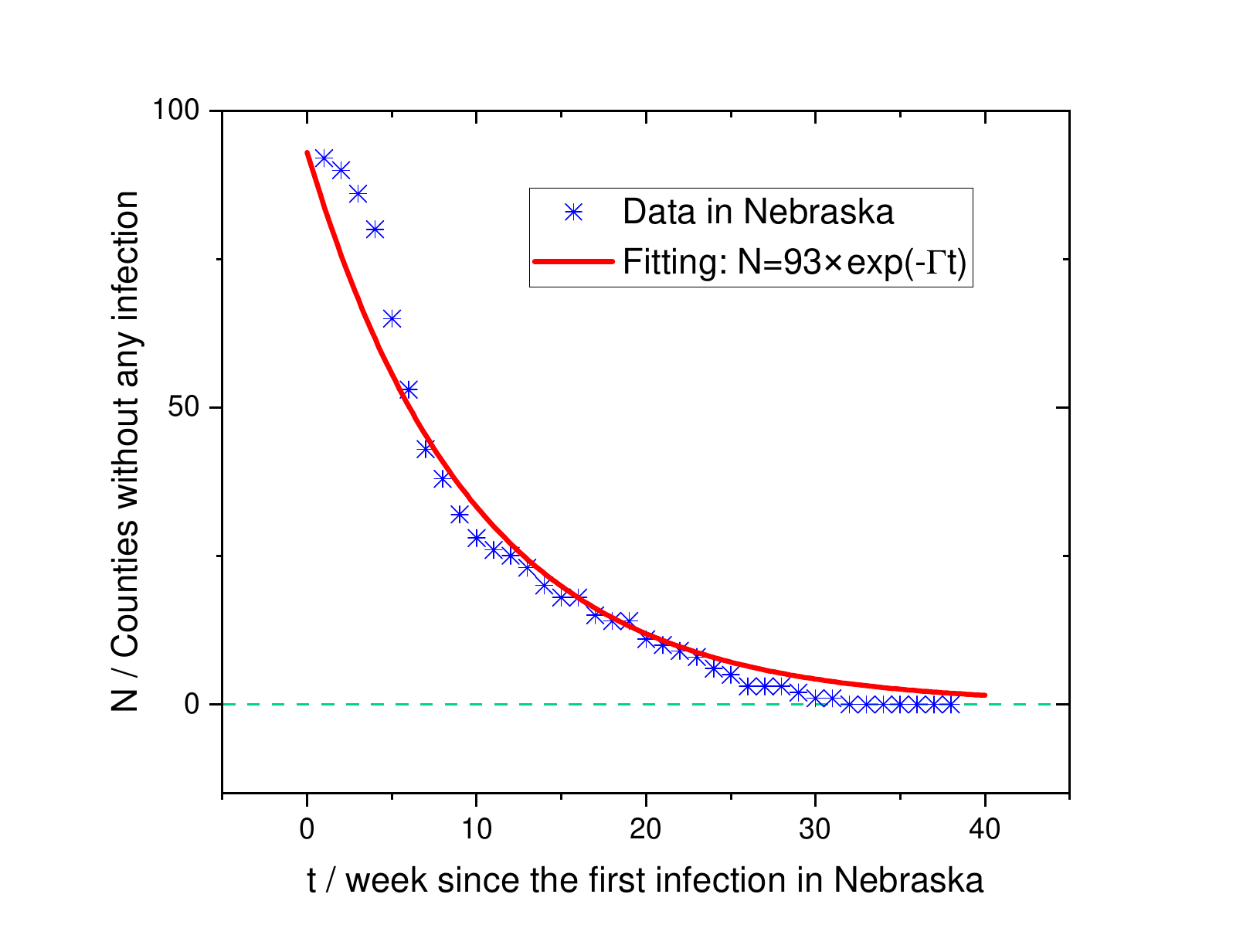}
  \caption{The number of counties with no infected citizens in Nebraska from 6th March 2020 to 27th November 2020. The red curve fits data points using single-exponential decay function $N=N_0 e^{-\Gamma t}$, where $N_0=93$ is the total number of counties in Nebraska. The infection rate is fitted as $\Gamma=0.1028$.}
  \label{fig:surviving-counties-Nebraska} 
\end{figure}

\begin{figure*}[th]
  \includegraphics[width = 2.0\columnwidth]{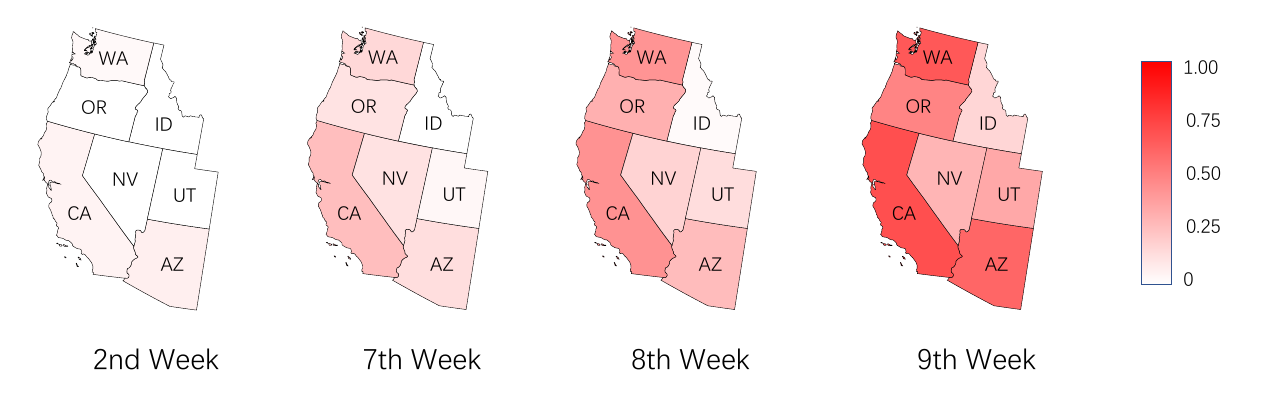}
  \caption{Illustration of infection spreading by weeks in the seven states: Arizona(AZ), California(CA), Idaho(ID), Nevada(NV), Oregon(OR), Utah(UT) and Washington(WA). The vividness of red for each state represents its proportion of counties having confirmed infected cases. The time record starts from 22nd January 2020, when Washington confirmed its first infected case of SARS-Cov-2.}
  \label{fig:geo-west-coast} 
\end{figure*}

In order to compare our simulations with infection spreading in the real world, we use the data provided by USAFacts (\url{https://usafacts.org/}), which has also been used in Ref.~\cite{Wong2020}. The data record the infected population by SARS-Cov-2 in every county of USA from 22nd January 2020 to 27th February 2022. Data is available from the Github repository \url{https://github.com/stccenter/COVID-19-Data/tree/master/US/County_level_summary}.

An individual of infected citizens is not a good choice as a single unit for checking our simulation. As shown in figure~\ref{fig:infected-population-King}, the infected population does not exhibit single-exponential decay behavior like the results in figure~\ref{fig:5-5-simulation-for-r0}. It is because the secondary transmission is omitted in our simulation. Thus, secondary transmission has to be involved when simulating infection spreading among individuals. To simulate it using the thermal dynamic model, one needs a time-dependent Hamiltonian, as explained in section~\ref{sec:thermal-dynamic-model}.

Figure~\ref{fig:surviving-counties-Nebraska} plots the number of counties  without any infected citizens in Nebraska from 6th March 2020 to 27th November 2020 by weeks. The data curve follows the deterministic SI model with secondary transmission~\cite{Lara2019AnalogyBT}, which has single-exponential decay behavior in the late time. The late time behavior is qualitatively consistent with the numerical results in figure~\ref{fig:infection-rate-system-bath-coupling}\textbf{a}. The disagreement between data and fitting in the early time is also due to the secondary transmission among counties, which should be involved to simulate this realistic scenario.

Figure~\ref{fig:geo-west-coast} illustrates the proportion of counties of seven states having confirmed infected cases in different weeks, since the first infected case in Washington state. The seven states are Arizona (AZ), California (CA), Idaho (ID), Nevada (NV), Oregon (OR), Utah (UT) and Washington (WA) in the USA. It can be compared with our simulation results in figure~\ref{fig:one-two-index-patients}. The pandemic was initially outbroken in WA, CA and AZ. The number of infected counties in the other states decays by distance from these three states. In particular, compare the infection spreading in OR and ID, we see that the spreading in OR is faster than that in ID. Because OR is adjacent to two infectious states WA and CA  while ID is adjacent to only WA. Similar phenomenon is also seen in our numerical simulation shown in figure~\ref{fig:one-two-index-patients}, where the infection rates is larger when introducing two patients more than one. On the other hand, difference between NV and UT seems not significant. It is because of the sparse and uneven population distribution in NV compared to that in UT. To simulate unevenly distributed population, one needs more nodes in the network than the evenly distributed population simulations. 

In summary, the comparison of the real-world data in figure~\ref{fig:surviving-counties-Nebraska} and the simulation results in figure~\ref{fig:infection-rate-system-bath-coupling} justifies that, the thermal dynamic model can well approximate the infection spreading processes among counties. Figure~\ref{fig:geo-west-coast} shows how geographical distance and population distribution influence infection spreading, as we have considered in the contact pattern of the simulation in figure~\ref{fig:one-two-index-patients}. Thus, equipped with the distance and population distribution information, one can predict the development of the pandemic in a network using the simulation approach, and guide us to introduce the intervention strategies at the proper time.

\section{Discussion}\label{sec:discussion}

In this work, we proposed a thermal dynamic model to simulate the infection spreading process in networks. Using perturbative analyses and numerical simulations, we proved that the quantum evolution of the thermal dynamic Hamiltonian simulates the Markovian process of the stochastic SI model. A systematic procedure is provided to determine the Hamiltonian parameters. As a demonstration, we simulated the infection spreading of the SARS-Cov-2 variant Omicron in a small-world network. We compared the simulation results with the real-world experimental data and showed their qualitative consistency.

This article provides a practical method of simulating Markovian processes on quantum computers. Due to the exponentially increasing dimension of Hilbert space of qubits, it is possible that more complicated Markovian processes can be simulated on quantum computers than that using classical algorithms, such as Monte-Carlo methods. Utilizing our proposed thermal dynamic model, we have seen that the stochastic SI model with long-distance transmission can be efficiently simulated. By tuning the inter-system coupling of the thermal dynamic Hamiltonian, complicated spreading processes, such as household and super-spreader transmission, can also be investigated. These processes are usually difficult to simulate using classical algorithms. 

The thermal dynamic Hamiltonian could have many variants to simulate more complicated infection spreading processes. In this work, we consider two compartments SI model simulated using the time-independent Hamiltonian. A more realistic SIR or SEIR model~\cite{Satorras_15}(R represents recovered or removed) with more compartments can be simulated by, for example, using more qubits to represent more status of an individual, and the removing/recovery processes can be simulated by resetting some of the infected individuals randomly to the removed state. Time-dependent Hamiltonian can be used to simulate the secondary transmission and the displacement of the population within communities. We leave more complicated simulations for further works.

We studied the infection spreading processes by simulating Markovian processes using quantum computers. This method can be applied to many other practical problems, as long as the simulations of Markovian processes are involved. For example, this method can be generalized to simulate the geometry Brownian motion used in finance for option pricing~\cite{Ross97,PhysRevE.105.L012106}, and many probabilistic forecasting problems in the areas of wind power~\cite{CARPINONE2015152} and solar irradiance~\cite{MUNKHAMMAR2019688}. On the other hand, the resetting operation used in our quantum evolution could also have some important applications, such as simulating diffusive particles, Lévy flights and fractional Brownian motion~\cite{PhysRevLett.106.160601,Majumdar_2018, Evans_2020}. Our work thus shed light on efficient simulations of these general models using quantum computers.

\section{Acknowledgements}
We thank Yunjun Zhang for the helpful introduction to the classical method of modeling epidemic processes. X.W. also thank Yusheng Gao for the illustration production.

\nocite{suzuki_1991,Feynman:1982,Sakurai2017,Fick1990,DORMAND198019}

\bibliographystyle{unsrt}
\bibliography{main.bib}

\end{document}


\newcommand{\PKU}{Center on Frontiers of Computing Studies, Peking University, Beijing 100871, China}

\newcommand{\PKUCS}{School of Computer Science, Peking University, Beijing 100871, China}

\title{Supplemental Material for ``Simulating the Spread of Infection in Networks with Quantum Computers"}

\maketitle

\section{Preliminaries on quantum simulation}\label{app:quantum-simulation}

In this section, we introduce the details of our numerical simulation and some mathematical notations used in the following sections.

Quantum simulation gets well known due to Richard Feynman~\mycite{36}{Feynman:1982}. It encodes a quantum system on several quantum bits (or qubits). Qubits are the basic building block of modern quantum computers. One qubit comprises a 2-dimensional Hilbert space that can be represented with computational basis $\ket{0}$ and $\ket{1}$, as we used in the main text. $N_q$ qubits comprise a $2^{N_q}$ dimensional Hilbert space following the principle of tensor product of many-body quantum systems. We denote this Hilbert space as $\mathcal{H}$. 

We briefly introduce how a quantum system evolves on a digital quantum computer with CNOT-based circuits. To encode a quantum system on qubits, we need to find a Hamiltonian $H$ to evolve a quantum state according to the Schr\"odinger equation
\begin{align}
    \ket{\psi (t)} =e^{-iHt} \ket{\psi (0)}.
\end{align}
Here we assume $H$ is independent on time. Due to the completeness of Pauli basis, the Hamiltonian can be written as a linear combination of multi-qubit Pauli operators
\begin{align}
    H = \sum_m h_m \sigma_m,
    \label{eq:general-hamiltonian}
\end{align}
where each $\sigma_m$ is a tensor product of single-qubit Pauli operators $\sigma_m=\sigma_{m_1}\ldots\sigma_{m_{N_q}},\sigma_{m_i}\in\{I,X,Y,Z\}$ and $h_m$ is a real number. However, digital quantum computers can not evolve this Hamiltonian directly. Instead, the evolution can be decomposed into several basic evolution blocks. The basic evolution blocks used in our numerical simulation are summarized below. Here we choose the basic gate as two-qubit $\mathrm{CNOT}$ gate~\mycite{19}{Nielsen2000}.
\begin{widetext}
\begin{equation}
\begin{aligned}
    \label{ReversedCircuit}
    \Qcircuit @C=1em @R=.7em {
     & \qw      & \multigate{2}{e^{-i\frac{\theta}{2}ZZ}}& \qw & & &\qw      & \ctrl{2}& \qw                              & \ctrl{2}&\qw   \\
     &          & \nghost{e^{-i\frac{\theta}{2}ZZ}}       &     & \push{\rule{.3em}{0em}=\rule{.3em}{0em}} & & &&&&  \\
     & \qw      & \ghost{e^{-i\frac{\theta}{2}ZZ}}       & \qw & & & \qw     & \targ         & \gate{e^{-i\frac{\theta}{2}Z}}   & \targ   &\qw 
    }
\end{aligned}
\end{equation}
\begin{equation}
\begin{aligned}
    \Qcircuit @C=1em @R=.7em {
     & \qw      & \multigate{2}{e^{-i\frac{\theta}{2}XX}}& \qw & & &\qw   & \gate{H}  & \ctrl{2}& \qw                              & \ctrl{2}&\gate{H}&\qw   \\
     &          & \nghost{e^{-i\frac{\theta}{2}XX}}       &     & \push{\rule{.3em}{0em}=\rule{.3em}{0em}} & & &&&&&&  \\
     & \qw      & \ghost{e^{-i\frac{\theta}{2}XX}}       & \qw & & & \qw  & \gate{H}  & \targ         & \gate{e^{-i\frac{\theta}{2}Z}}   & \targ &\gate{H} &\qw 
    }
\end{aligned}
\end{equation}
\end{widetext}

With these evolution blocks, the Schr\"odinger evolution can be carried out using the first-order Trotter decomposition  (or higher order Suzuki-Trotter decomposition~\mycite{35}{suzuki_1991} for higher accuracy)
\begin{align}\label{eq:Trotter-error-equation}
    e^{-iHt} = (\prod_{m}e^{-i\delta t h_m\sigma_m})^{N}+\mathcal{O} (t^2/N),
\end{align}
where $N$ is the total number of Trotter steps to evolve for time $t$, $\delta t\equiv t/N$ denotes the time length of one Trotter step. The systematic error due to Trotter decomposition can be controlled by increasing $N$ for a given evolution time $t$. In our numerical simulations, we choose Trotter step $\delta t=0.01$ if not specified.

The Schr\"odinger equation describes an isolated system's evolution. For a non-isolated system's evolution, the description is more complicated. Specifically, we consider the system accompanied by a heat bath. Their state as a whole is described by density operator $\rho_{sb}$, which is a non-negative, trace-1 linear operator~\mycite{37}{Sakurai2017}. The system is also described by density operator $\rho_s$ in Hilbert space $\mathcal{H}$, which is given by
\begin{align}\label{eq:partial_trace}
    \rho_s=\Tr_b (\rho_{sb}),
\end{align}
where $\Tr_b$ is the partial trace operation over the bath Hilbert space. 

Assume the whole state $\rho_{sb}$ still follows the Schr\"odinger equation. The system's evolution alone is a thermal dynamic evolution, which can be regarded as a map from one density operator to another. Generally, such a map is a completely positive and trace preserving (CPTP) map that can be represented by
\begin{align}
    S (\rho_s) = \sum_i A^i \rho_s A^{i\dagger}.
    \label{eq:quantum-channel}
\end{align}
We also call this map a quantum channel. As a quantum channel is also a linear map on the space of density operator, we can simplify this representation by introducing \emph{Hilbert-Schmidt space} denoted by $\Bcal (\mathcal{H})$, where each density operator is represented by a $2^{2N}$ column vector 
\begin{align}
    \rho_s \rightarrow \supket{\rho_s}.
\end{align}

Hilbert-Schmidt space is equipped with an inner product $\supbraket{\rho_1}{\rho_2}=\Tr (\rho_1^{\dagger}\rho_2)$, which possesses the physical meaning of the possibility of measuring state $\rho_1$ given state $\rho_2$. We call the column vector $\supket{B}$ \emph{superket} and the dual vector $\supbra{A}$ \emph{superbra}. The superket can be expanded with an orthogonal basis which is induced by an orthogonal basis $\{\ket{m}\}$ (like the energy eigenstates) in Hilbert space $\mathcal{H}$
\begin{align}
     \supket{\phi_{nm}^{ (0)}}\equiv \supket{\ket{n}\bra{m}}.
     \label{eq:supket-basis}
\end{align}
The meaning of superscript $ (0)$ will get clear later in section~\ref{app:evolution-expansion}. Thus Eq.~(\ref{eq:quantum-channel}) can be represented by
\begin{equation}
\begin{aligned}
     \supbra{\phi_{nm}^{ (0)}} S\supket{\rho_s} &= \sum_{kl}S_{nm;kl}(\rho_s)_{kl},\\
    S_{nm;kl}&\equiv\sum_i A^i_{nk}A^{i*}_{ml},
\end{aligned}
\end{equation}
where $\rho_{s\;kl}=\bra{k}\rho_s\ket{l},A^i_{nk}=\bra{n}A^i\ket{k}$ and $*$ denote complex conjugate. Here we use the definition of inner product in Hilbert-Schmidt space, i.e., $\supbra{\phi_{nm}^{ (0)}} S\supket{\rho_s}=\Tr (\ket{m}\bra{n}S (\rho_s))=\bra{n}S (\rho_s)\ket{m}$.

Another key element in our simulation is resetting operation, which can also be described using the language of density operator and partial trace. Consider a system composited by subsystems A and B. It is described by a density operator $\rho$. If system A is reset to a pure state $\ket{k'_A}$, the whole system's density operator becomes
\begin{equation}
\begin{split}
    \rho'=\ket{k'_A}\bra{k'_A}\otimes\rho^B
    \label{eq:partial_trace_outcome}
\end{split}
\end{equation}
where $\rho^B=\Tr_A{\rho}$ is the reduced density operator on the Hilbert space of the subsystem B.

Resetting can be realized in a quantum circuit if a perfect measurement of a subsystem is possible. Assume $\{\ket{k_A}\}$ is a complete basis in the Hilbert space of system A. Performing a projection measurement $P_k=\ket{k_A}\bra{k_A}\otimes I_B$ on the initial density operator $\rho$, the outcome state is 
\begin{equation}
    \rho (k)=\ket{k_A}\bra{k_A}\otimes\rho^B,
\end{equation}
with the probability
\begin{equation}
    p (k)=\Tr{\rho^A\ket{k_A}\bra{k_A}},
\end{equation}
where $\rho^A=\Tr_B{\rho}$ is the reduced density operator on the Hilbert space of system A. After the measurement, according to the measurement result, we transform the outcome state $\ket{k_A}$ unitarily to $\ket{k'_A}$ using the quantum gate while keeping $\rho^B$ unchanged. The final state is
\begin{equation}
    \rho'=\ket{k'_A}\bra{k'_A}\otimes\rho^B,
\end{equation}
which is the same as the partial trace outcome \eqref{eq:partial_trace_outcome}.

A typical example is when subsystem A is a single qubit and $\ket{k'_A}=\ket{0}_A$. The resetting operation of this case can be represented using a quantum circuit
\begin{align}
    \label{Reset_circuit}
  \Qcircuit @C=1em @R=.7em {
\lstick{A}& \meter & \qw & \gate{\ket{0}} & \qw \\
\lstick{B}& \qw & \qw & \qw & \qw
}
\end{align}
where the first gate is a measurement in $\{\ket{0},\ket{1}\}$ basis, and the second gate is an X-gate if the measurement outcome is $\ket{1}$ while doing nothing if the outcome is $\ket{0}$.

\section{Perturbative evolution expansion}\label{app:evolution-expansion}
In this section, we briefly review the general formalism of the time evolution of a system with a weakly coupled heat bath that has been studied in Ref.~\mycite{22}{Terhal2000}. The perturbative analysis of the evolution helps to build the relationship between the classical infection-recovery Markovian processes and quantum evolution, as will be discussed in the next section. The Hamiltonian of the whole system reads
\begin{align}
    H = H_{s}\otimes \Id_{bath}+\Id_{sys}\otimes H_{b} +\lambda H_{sb}.
    \label{eq:total-Hamiltonian}
\end{align}
In our simulation, the coupling term $H_{sb}$ can be formally written as 
\begin{align}
    H_{sb} = \sum_i S^i\otimes B^i.
    \label{eq:coupling-Hamiltonian}
\end{align}
The whole system follows the exact Schr\"odinger evolution
\begin{align}
    \rho_s\otimes\rho_b \rightarrow e^{-iHt}\rho_s\otimes\rho_{b,\beta} e^{iHt},
    \label{eq:total-evolution}
\end{align}
where $\rho_s$ and $\rho_{b,\beta}$ are the system's and bath's initial density operators, respectively. $\beta=1/(k_B T)$ is the inverse temperature of the heat bath. The heat bath is set to be equilibrium state
\begin{align}
    \rho_{b,\beta}=\frac{e^{-\beta H_{b}}}{\Tr (e^{-\beta H_{b}})}.
    \label{eq:bath-equilibrium-state}
\end{align}
Theoretically, thermal evolution requires an infinitely large heat bath, which requires an infinitely large number of qubits. To reduce the computational resource, we couple each system qubit with one bath qubit and reset it to its equilibrium state once after a specific time interval $\Delta t_b$. In our numerical simulation, the time interval is equal to the time interval of resetting the infectious node to $\ket{1}$, i.e., $\Delta t_b=\Delta t$. The resetting operation is a good approximation to mimic an infinitely large heat bath. In our simulation, since we take $\beta\rightarrow\infty$, the equilibrium state reduces to the ground state of $H_{b}$ according to Eq.~(\ref{eq:bath-equilibrium-state}).

We aim to analyze how systems evolve under the evolution Eq.~(\ref{eq:total-evolution}). The evolution of the system density operator is derived by tracing out the bath Hilbert space
\begin{align}
    S_{\lambda,t} (\rho_s) \equiv  \Tr_{b} (e^{-iHt}\rho_s\otimes\rho_{b,\beta} e^{iHt}),
    \label{eq:system-density-evolution}
\end{align}
where $S_{\lambda,t}$ is a CPTP map acting on the initial system density operator $\rho_s$. In Ref.~\mycite{22}{Terhal2000}, the authors show that this CPTP map can be treated perturbatively with the small parameter $\lambda$. The map can be expanded by 
\begin{align}
    S_{\lambda,t} = S^{ (0)}_{t}+\lambda S^{ (1)}_{t}+\lambda^2 S^{ (2)}_{t}+\lambda^3 S^{ (3)}_{t}+\mathcal{O} (\lambda^4).
    \label{eq:evolution-expansion}
\end{align}
where $S^{ (0)}_{t}$ is the evolution of the system alone without the coupling of the heat bath, i.e., $S^{ (0)}_{t}=S_{\lambda=0,t}$. In that case, the system's evolution can be solved exactly by observing
\begin{equation}
\begin{aligned}
    S^{ (0)}_{t}\supket{\phi_{nm}^{ (0)}}&= e^{-iH_{s}t}\ket{n}\bra{m}e^{iH_{s}t} \\
    &=e^{-i (E_n-E_m)t}\supket{\phi_{nm}^{ (0)}},
     \label{eq:zero-order-evolution}
\end{aligned}
\end{equation}
where $\ket{n}$ is the eigenstate of the system's Hamiltonian with eigenvalue $E_n$  satisfying 
\begin{align}
    H_{s}\ket{m} = E_m\ket{m}.
\end{align}

\subsection*{Classical and purely quantum evolution}

In this subsection, we show that using the thermal dynamic Hamiltonian Eq.~(\ref{eq:total-Hamiltonian}), a qubit would behave like a classical spin. Specifically, we prove that in case: (1) The system-bath coupling $\lambda$ is small. (2) The initial state is the energy eigenstate of the system. (3) The measured operator is a linear combination of the projectors onto the system's energy eigenstates; then, the measured expectation values will be dominated by the classical part of the evolution. Note that due to the linearity of the quantum channel, the latter two conditions are reduced to that the initial state and the measured operators are all like $\ket{n}\bra{n}=\supket{\phi_{nn}^{ (0)}}$.

First, we distinguish between evolution's classical and purely quantum parts. They can be distinguished explicitly in the zeroth order of the evolution expansion  (\ref{eq:evolution-expansion}). Note that $S^{ (0)}_{t}$ is diagonal under the basis $\supket{\phi_{nm}^{ (0)}}$ according to Eq.~(\ref{eq:zero-order-evolution})
\begin{equation}
\begin{aligned}
    S^{ (0)}_{t} &= \sum_{mn} e^{-i (E_n-E_m)t}\supket{\phi_{nm}^{ (0)}}\supbra{\phi_{nm}^{ (0)}}\\
    &=S_{t}^{c (0)}+S_{t}^{q (0)},
\end{aligned}
\end{equation}
with
\begin{align}
 \left\{ \begin{array}{ll}
 S_{t}^{c (0)} & \equiv \sum_n\supket{\phi_{n}^{ (0)}}\supbra{\phi_{n}^{ (0)}}\\
 S_{t}^{q (0)} & \equiv \sum_{n\neq m} \mu_{nm}^{ (0)}\supket{\phi_{nm}^{ (0)}}\supbra{\phi_{nm}^{ (0)}},
  \end{array} \right.
  \label{eq:zero-order-form}
\end{align}
where we denote $\supket{\phi_{n}^{ (0)}}\equiv\supket{\phi_{nn}^{ (0)}}$ and $\mu_{mn}^{ (0)}\equiv e^{-i (E_n-E_m)t}$. If we assume a non-degenerate energy spectrum of the system, the phase accumulation $e^{-i (E_n-E_m)t}$ in $S_{t}^{q (0)}$ can be regarded as a purely quantum effect. It will vanish under an infinitely long time average. The classical part $S_{t}^{c (0)}$ leaves the energy eigenstates $\ket{n}\bra{n}=\supket{\phi^{ (0)}_n}$ unchanged. If we treat each $\supket{\phi^{ (0)}_n}$ as a classical state, $S_{t}^{c (0)}$ can also be regarded as a classical trivial stochastic matrix under the energy eigenbasis, satisfying normalisation condition
\begin{align}
    \sum_l \supbra{\phi_l^{ (0)}} S_{t}^{c (0)} \supket{\phi_k^{ (0)}}=1,\; \forall k.
\end{align}

Apply the above analysis to the infection spreading processes. We choose the system Hamiltonian 
\begin{align}
    H_{s} = -\sum_{i\in I}\sum_{j\in S}\gamma_{ij}Z^s_i Z^s_j,
    \label{eq:system-hamiltonian}
\end{align}
of which eigenstates are tensor products of computational basis $\ket{0}$ and $\ket{1}$. These eigenstates are the initial states as well as the final states in which we are interested. A classical trivial stochastic matrix $S_{t}^{c (0)}$ leaves an input eigenstate unchanged. Thus, population does not diffuse if the coupling between system and bath vanishes.

Consider the higher order correction of $S_{\lambda,t}$. As shown in the next section, if we take the explicit form of the thermal dynamic Hamiltonian, the odd-order corrections $S_t^{ (2k+1)}$ of $S_{\lambda,t}$ vanish. Thus the leading order  $S_t^{ (1)}$ have no contribution to the thermal evolution and the next leading order $S_t^{ (2)}$ will bring out correction to eigenvalues and eigenvectors in Eq.~(\ref{eq:zero-order-form}). The correction can be evaluated similarly to the formal perturbative correction to the eigenvalues and eigenvectors for Hermitian Hamiltonian, as introduced in many quantum mechanics textbook~\mycite{37}{Sakurai2017}. One needs to notice that as the eigenvalues in $S_{t}^{c (0)}$ are degenerate while in $S_{t}^{q (0)}$ are not, we need to carry out degenerate perturbation and non-degenerate one respectively on these two parts. The degenerate perturbation selects the true bases in the degenerate subspace by solving a secular equation regarding $S^{ (2)}_{t}$. The selected true bases are superpositions of the original bases
\begin{align}
    \supket{\tilde{\phi}_m^{ (0)}}\equiv \sum_n V_{mn} \supket{\phi_n^{ (0)}}.
\end{align}
where $V$ represents a unitary transformation. Consider the second-order correction to the Eq.~(\ref{eq:zero-order-form}). The result reads
\begin{widetext}
\begin{align}
    &\left\{ \begin{array}{ll}
 S_{\lambda,t}^c & =\sum_n (1+\lambda^2 S^{ (2)}_{t\; \tilde{n};\tilde{n}}+\mathcal{O} (\lambda^4))\left[\supket{\tilde{\phi}_{n}^{ (0)}}\supbra{\tilde{\phi}_{n}^{ (0)}}+\mathcal{O} (\lambda^2)\right],\\
 S_{\lambda,t}^q & =\sum_{n\neq m}  (\mu_{nm}^{ (0)}+\lambda^2 S^{ (2)}_{t\; nm;nm}+\mathcal{O} (\lambda^4))\supket{\phi_{nm}}\supbra{\phi_{nm}},
  \end{array} \right.
\end{align}
where 
\begin{equation}
\begin{aligned}
  \supket{\phi_{nm}} \equiv \supket{\phi_{nm}^{ (0)}}+\lambda^2\sum_l \supket{\tilde{\phi}_l^{ (0)}}\frac{S^{ (2)}_{t\; \tilde{l};nm}}{\mu_{nm}^{ (0)}-1}
  +\lambda^2\sum_{\substack{n'\neq m'\\n'm'\neq nm}} \supket{\phi_{n'm'}^{ (0)}}\frac{S^{ (2)}_{t\; n'm';nm}}{\mu_{nm}^{ (0)}-\mu_{n'm'}^{ (0)}}
  +\mathcal{O} (\lambda^4),\\
  S^{ (2)}_{t\; \tilde{n};\tilde{n}}\equiv\supbra{\tilde{\phi}_{n}^{ (0)}}S^{ (2)}_t\supket{\tilde{\phi}_{n}^{ (0)}},\quad 
  S^{ (2)}_{t\; \tilde{l};nm}\equiv \supbra{\tilde{\phi}_{l}^{ (0)}}S^{ (2)}_t\supket{\phi_{nm}^{ (0)}},\quad
  S^{ (2)}_{t\; n'm';nm}\equiv \supbra{\phi_{n'm'}^{ (0)}}S^{ (2)}_t\supket{\phi_{nm}^{ (0)}}.
  \label{eq:second-order-form}
\end{aligned}
\end{equation}
\end{widetext}
In a classical stochastic process, the input and output states are all classical states, denoted by  $\supket{\phi_k^{ (0)}}$ and $\supket{\phi_l^{ (0)}}$ respectively and $l\neq k$. Thus after the quantum evolution for time $t$ under system-bath coupling $\lambda$, the probability of measuring $\supket{\phi_l^{ (0)}}$ receiving contributions from the classical part and the purely quantum part at the leading order is
\begin{align}
    \left\{ \begin{array}{ll}
 \supbra{\phi_l^{ (0)}}S_{\lambda,t}^c \supket{\phi_k^{ (0)}}& =\mathcal{O} (\lambda^2),\\
 \supbra{\phi_l^{ (0)}}S_{\lambda,t}^q \supket{\phi_k^{ (0)}}& =\mathcal{O} (\lambda^4).
  \end{array} \right.
\end{align}
We find $\supbra{\phi_l^{ (0)}}S_{\lambda,t}^q  \supket{\phi_k^{ (0)}}\ll \supbra{\phi_l^{ (0)}}S_{\lambda,t}^c \supket{\phi_k^{ (0)}}$ for a small enough $\lambda$. 

The above perturbative analysis shows that the classical evolution is dominant during the evolution. In other words, the system's density operator can be regarded as a classical distribution. This is the reason that we say a qubit behaves like a classical spin in the main text. The $\mathcal{O} (\lambda^2)$ dependence of the evolution has been explicitly shown in our numerical simulation (see Fig. 2\textbf{b} in the main text). In the next section, we will focus on calculating the classical part of the evolution.

\section{Liouvillian formalism and calculation for evolution expansion}\label{app:Liouvillian-formalism-and-calculation-for-evolution-expansion}
To find an explicit form of the evolution expansion in Eq.~(\ref{eq:evolution-expansion}), we introduce the Liouvillian formalism~\mycite{38}{Fick1990} for the time evolution of the density operator. 

In Schr\"odinger picture, the time evolution of the density operator is given by the von-Neumann equation
\begin{align}
    \frac{\ud \rho (t)}{\ud t}=-i[H,\rho (t)]\equiv -iL\rho (t).
    \label{eq:von-Neumann}
\end{align}
Here  $L$ is called the Liouville operator, the linear Lie product operator of the system Hamiltonian $H$. If the Hamiltonian is explicitly time-independent, this equation can be integrated as 
\begin{align}
    \rho (t) = e^{-iLt}\rho (0),
\end{align}
where $e^{-iLt}$ is defined by its Taylor expansion
\begin{align}
    e^{-iL t}\rho (0)=\sum_k \frac{1}{k!} (-it)^k L^k\rho=e^{-iHt}\rho e^{iHt}.
\end{align}
Thus we recovered the time evolution of the density operator as we used in Eq.~(\ref{eq:system-density-evolution}). The evolution operator on the system is given by 
\begin{align}
    S_{\lambda,t} (\rho_s) \equiv  \Tr_{b} (e^{-iLt}\rho_s\otimes\rho_{b,\beta}).
    \label{eq:system-evolution}
\end{align}

To find an explicit form of $e^{-iL t}$, we first decompose the Liouville operator of the system-bath Hamiltonian into two parts
\begin{align}
    L=L_1+L_2; L_1\equiv (L_s+L_b), L_2\equiv\lambda L_{sb},
\end{align}
where $L_s,L_b$ and $L_{sb}$ are Liouville operator of $H_{s},H_{b}$ and $H_{sb}$ respectively. The operator identity~\mycite{38}{Fick1990} for time-independent $L_1,L_2$ reads
\begin{align}
    e^{-iLt}=e^{-iL_1 t}-i\int_0^t \ud t' e^{-iL_1  (t-t')}L_2e^{-iLt'}\ud t'.
\end{align}
The desired evolution $e^{-iLt}$ is shown on both sides of this identity. Since $\lambda$ in $L_2$ is a small quantity, iteratively taking the left hand side $e^{-iLt}$ into the right hand side  $e^{-iLt'}$ gives the perturbative expansion of $e^{-iLt}$. Taking this expansion into Eq.~(\ref{eq:system-evolution}), we find the pertubative formula of $S_{\lambda,t}$ for each order of $\lambda$. For example, the first-order expansion reads
\begin{equation}
\begin{aligned}
     S^{ (1)}_{t} (\rho_s)=\Tr_b (&-i\int_0^{t}\ud t_1  e^{-i (L_s+L_b) (t-t_1)}\\
     &L_{sb} e^{-i (L_s+L_b)t_1} \rho_s\otimes\rho_{b,\beta}).
\end{aligned}
\end{equation}

We show that the odd order terms vanish if we take thermal dynamic Hamiltonian as in Eq.~(\ref{eq:total-Hamiltonian}), where
\begin{align}
    H_s =& -\sum_{i\in I}\sum_{j\in S} \gamma_{ij}Z^s_i Z^s_j\nonumber \\
     H_{sb}=   & - \sum_{i\in I\cup S} X^s_i X^b_i\nonumber \\
     H_b=   & - \alpha \sum_{i\in I}\sum_{j\in S} Z^b_i Z^b_j.
        \label{eq:main-hamiltonian}
\end{align}
It is the Hamiltonian Eq. (24) used in the main text. Take the first order term as an example. Because $\rho_{b,\beta}$ is the equilibrium state of $H_b$, it commutes with $H_b$ and $e^{-iL_b (t_1)}\rho_{b,\beta}=\rho_{b,\beta}$. Then, we take system-bath coupling Hamiltonian as in Eq.~(\ref{eq:coupling-Hamiltonian}). It leads to 
\begin{align}
    S^{ (1)}_{t} (\rho_s)=-i\Tr_b\left (\int_0^t \ud t_1~ e^{-i (L_s+L_b) (t-t_1)}\right.\nonumber\\
    \left. \sum_i[S^i \otimes B^i,\rho_s (t_1)\otimes \rho_{b,\beta}]\right),
\end{align}
where $\rho_s (t_1)$ is the time evolved $\rho_s$
\begin{align}
    \rho_s (t_1)=e^{-iH_s t_1}\rho_s e^{iH_s t_1}.
\end{align}
Expand the commutator and use the cyclic permutation invariance of trace. We have
\begin{align}
    S^{ (1)}_{t} (\rho_s)=&-i\sum_i \int_0^t \ud t_1~ e^{-iL_s (t-t_1)}\nonumber\\
    &\left (S^i\rho_s (t_1)-\rho_s (t_1)S^i\right) \Tr_b (B^i \rho_{b,\beta}).
    \label{eq:first-order-correction}
\end{align}
This equation is a generalization of Eq.~(2.72) in Ref.~\mycite{22}{Terhal2000}. We take $B^i$ as $X_i^b$ and $H_b$ as in Eq.~(\ref{eq:main-hamiltonian}). We see this equation vanishes by expanding the trace with $H_b$'s eigenstates  $\ket{n}$
\begin{align}
    \Tr_b (B^i \rho_{b,\beta})=\sum_n e^{-\beta E_n}\langle n|X_i|n\rangle=0,
\end{align}
where the last equality hold since $|n\rangle$ takes all possible configurations of classical bit string like $|1011\ldots \rangle$. 

Similarly, The third order correction $S^{ (3)}_{t} (\rho_s)$ has terms with coefficients like
\begin{align}
    \Tr_b (B^k (t')B^j (t'')B^i (t''')\rho_{b,\beta}),
\end{align}
where $t',t'',t'''$ denote some evolution times. The trace can be expanded similarly
\begin{align}
    \sum_{nml} e^{-iE_n (t'-t'''-i\beta)}e^{-iE_m (t''-t')}e^{-iE_l (t'''-t'')}\nonumber\\
    \melem{n}{X_k}{m}\melem{m}{X_j}{l}\melem{l}{X_i}{n}=0.
\end{align}
Thus the third order correction vanishes since any bit strings can not be recovered by flipping ($0\leftrightarrow 1$ exchange) odd times. We can generalize the result to arbitrary odd order corrections, which means $S_t^{ (2k+1)} (\rho_s)$ vanishes for all non-negative integer $k$. On the other hand, only the even order terms need to be considered.

The lowest order term $S_t^{ (2)}$ can be calculated following a similar procedure, as given in \mycite{22}{Terhal2000}. We only focus on the classical part of the evolution, which means that we are interested in the matrix elements of $S_t^{ (2)}$ between two classical states $\supbra{\phi_m^{ (0)}}S_t^{ (2)} \supket{\phi_n^{ (0)}}$. The matrix element has a physical meaning of the probability of observing the system's final state $|m\rangle$, given an initial state $|n\rangle$. The matrix element can be evaluated exactly. The result is 
\begin{widetext}
\begin{equation}
\begin{aligned}
    \supbra{\phi_m^{ (0)}}S_t^{ (2)} \supket{\phi_n^{ (0)}}=2\int_{-\infty}^{+\infty} \frac{\ud \omega}{2\pi} \sum_{ij}\tilde{h}^{ij} (\omega) &\left[S^j_{mn}S^i_{nm}\frac{1-\cos ( (\omega+E_n-E_m)t)}{ (\omega+E_n-E_m)^2}\right.\\ &\left.-\delta_{mn}\sum_l\frac{S^i_{ml}S^j_{ln} (1-\cos ( (\omega+E_n-E_l)t))}{ (\omega+E_n-E_l)^2}\right],
    \label{eq:second-order-evolution}
\end{aligned}
\end{equation}
\end{widetext}
where $S^j_{mn}=\langle m|S^j|n\rangle$, and $\tilde{h}^{ij} (\omega)$ is defined as a spectrum function of the heat bath
\begin{equation}
\begin{aligned}
    \tilde{h}^{ij} (\omega)=&\int_{-\infty}^{+\infty} \ud t~ e^{-i\omega t}h^{ij} (t),\\
    h^{ij} (t)\equiv& \Tr_b (B^i e^{-iH_b t}B^j e^{iH_bt} \rho_{b,\beta}).
    \label{eq:bath-spectrum}
\end{aligned}
\end{equation}

\subsection{Application to the thermal dynamic Hamiltonian}

We apply the above perturbative results Eq.~(\ref{eq:second-order-evolution}) to the thermal dynamic Hamiltonian. First, we calculate the spectrum of the heat bath taking bath Hamiltonian $H_b$ in Eq.~(\ref{eq:main-hamiltonian}) and $B^i=X^b_i$. With the commutation relations of Pauli operators, it gives
\begin{align}
      h^{ij} (t)=\left\{ \begin{array}{ll}
 \exp{-2i\sum_{i'\in I}\alpha_{i'j}t},&i=j\in S,\\
 \exp{-2i\sum_{j'\in S}\alpha_{ij'}t},&i=j\in I,\\
 0, &i\neq j.
  \end{array} \right.
\end{align}
Here we take the zero temperature limit $\beta\rightarrow\infty$ so that only the ground states of the heat bath contribute, i.e., the spin parallel states. The spectrum function can be derived by Fourier transformation according to Eq.~(\ref{eq:bath-spectrum})
\begin{equation}
\begin{aligned}
     \tilde{h}^{ij} (\omega)=\left\{ \begin{array}{ll}
 2\pi \delta (\omega+2\sum_{i'\in I}\alpha_{i'j}),&i=j\in S,\\
 2\pi \delta (\omega+2\sum_{j'\in S}\alpha_{ij'}),&i=j\in I,\\
 0,& i\neq j.
  \end{array} \right.
  \label{eq:spectrum-function}
\end{aligned}
\end{equation}
Thus for the matrix elements of $S_t^{ (2)}$, only terms with identical $i,j$ need to be considered. Then take $S^i=X^s_i$. One finds $\supbra{\phi_m^{ (0)}}S_t^{ (2)} \supket{\phi_n^{ (0)}}$ does not vanish only if  (1)~$m=n$, where the second term of Eq.~(\ref{eq:second-order-evolution}) contributes.  (2)~$m,n$ differ at one spin site, where the first term of Eq.~(\ref{eq:second-order-evolution}) contributes. Also due to $S^i=X^s_i$, the summation $\sum_l$ in Eq.~(\ref{eq:second-order-evolution}) only takes the configuration $l$ that differs from $m (n)$ at one spin site. With some algebra, the diagonal matrix elements of $S_t^{ (2)}$ read
\begin{align}
    \supbra{\phi_n^{ (0)}}S_t^{ (2)} \supket{\phi_n^{ (0)}}= -2\left[\sum_{j\in S} A_{n,j} (t)+\sum_{i\in I} B_{n,i} (t)\right],
    \label{eq:diagonal-matrix-element}
\end{align}
where 
\begin{equation}
\begin{aligned}
    A_{n,j} (t)&=\frac{1-\cos (-2\sum_{i'\in I}\alpha_{i'j}+E_n-E_{X_j (n)})t}{ (-2\sum_{i'\in I}\alpha_{i'j}+E_n-E_{X_j (n)})^2},\\
    B_{n,i} (t)&=\frac{1-\cos (-2\sum_{j'\in S}\alpha_{ij'}+E_n-E_{X_i (n)})t}{ (-2\sum_{j'\in S}\alpha_{ij'}+E_n-E_{X_i (n)})^2}.
    \label{eq:Anj-Bnj}
\end{aligned}
\end{equation}
Here we use $X_j (n)$ to represent the bit string $n$ flipped at node $j$. The off-diagonal elements can be derived similarly
\begin{equation}
\begin{aligned}
     \supbra{\phi_{X_j  (n)}^{ (0)}}S_t^{ (2)} \supket{\phi_n^{ (0)}}&=2 A_{n,j} (t),\quad j\in S,\\
     \supbra{\phi_{X_i  (n)}^{ (0)}}S_t^{ (2)} \supket{\phi_n^{ (0)}}&=2 B_{n,i} (t),\quad i\in I,
     \label{eq:off-diagonal-matrix-element}
\end{aligned}
\end{equation}
while the other matrix elements vanish. According to the above two equations, one finds the relationship 
\begin{align}
    \sum_m \supbra{\phi_{m}^{ (0)}}S_t^{ (2)} \supket{\phi_n^{ (0)}}=0.
    \label{eq:normalization-condition}
\end{align} 
It indicates that each column of the evolution matrix $S_{\lambda,t}\simeq1+\lambda^2 S_t^{ (2)}$ summed to $1$, thus $S_{\lambda,t}$ retains a stochastic matrix up to $\mathcal{O}(\lambda^2)$.

Another essential element in the Hamiltonian simulation is resetting infectious nodes to $\ket{1}$ once after having unitary evolution $S_{\lambda,\Delta t}$ with a short time interval $\Delta t$. We first demonstrate how it behaves in the classical part of the evolution and then prove the following property of the thermal dynamic evolution. This property allow us to pre-determine the inter-bath coupling $\alpha$ as discussed in section III.D in the main text. 

\textit{Property}. The infection rate of susceptible node $j$ tends to $0$ in the limit $\gamma_{ij}\rightarrow 0,\forall i\in I$ up to $\mathcal{O} (\lambda^2)$, if the system-bath coupling $\sum_{i'\in I}\alpha_{i'j}\Delta t=k\pi,k=1,2,\ldots$. 

\textit{Proof.} Given a classical probability distribution $\{p_{m^S,m^I}\}$, the spin configuration reads
\begin{align}
    \supket{\psi}&\equiv \sum_{m^S,m^I}p_{m^S,m^I}\supket{\phi^{ (0)}_m}\\
    &=\sum_{m^S,m^I}p_{m^S,m^I}\supket{m_1^Sm_2^S\ldots m_{|S|}^Sm_1^I m_2^I\ldots m_{|I|}^I},
    \label{eq:classical-spin-configuration}
\end{align}
where each $m_i^{S (I)}$ taking values $0$ or $1$. Susceptible nodes and infectious nodes are distinguished explicitly in the second line. The resetting operation $\mathbf{R}$ on the infectious set transforms it into
\begin{align}
    \mathbf{R}\supket{\psi}=\sum_{m^S} (\sum_{m^I}p_{m^S,m^I})\supket{m_1^S m_2^S\ldots m_{|S|}^S 11\ldots 1}.
    \label{eq:reset-action}
\end{align}
Here $\sum_{m^I}p_{m^S,m^I}$ is the probability of measuring the corresponding final state. Thus $\mathbf{R}$ is also a stochastic matrix with unit entries at row $\supket{m_1^S m_2^S\ldots m_{|S|}^S 11\ldots 1}$ and the corresponding columns $\supket{m_1^S m_2^S\ldots m_{|S|}^S m_1^Im_2^I\ldots m_{|I|}^I}$.

To prove the property of zero infection rate  up to $\mathcal{O} (\lambda^2)$, first notice that if we set $\gamma_{ij}\rightarrow 0$, the energy differences $E_n-E_{X_j (n)}$ will all go to zero. Then taking $\sum_{i'\in I}\alpha_{i'j}\Delta t=k\pi$, we find $A_{n,j} (\Delta t)$ vanish as indicated in Eq.~(\ref{eq:Anj-Bnj}). If we further prove that $B_{n,j}$ has no contribution in the whole evolution involving resetting operation, then $S_{\Delta t}^{ (2)}$ is effectively zero. Thus the system does not evolve up to $\mathcal{O} (\lambda^2)$, and the infection rate is approximately zero. Consider the adjoint action $\mathbf{R} S_{\lambda,\Delta t}$. The n-th column of $S_{\lambda,\Delta t}$
\begin{equation}
\begin{aligned}
    S_{\lambda,\Delta t} \supket{\phi_n^{ (0)}}&= \sum_m \supket{\phi_{m}^{ (0)}}\supbra{\phi_{m}^{ (0)}}S_{\lambda,\Delta t} \supket{\phi_n^{ (0)}}\\
    &=\sum_{m^S,m^I} \supket{\phi_{m^S,m^I}^{ (0)}}\supbra{\phi_{m^S,m^I}^{ (0)}}S_{\lambda,\Delta t} \supket{\phi_n^{ (0)}}
    \label{eq:S-on-phi-n}
\end{aligned}
\end{equation}
can be regarded as a classical probability distribution as in  (\ref{eq:classical-spin-configuration}). The normalization condition
\begin{align}
    \sum_{m^S,m^I}\supbra{\phi_{m^S,m^I}^{ (0)}}S_{\lambda,\Delta t} \supket{\phi_n^{ (0)}}=1
\end{align}
holds according to Eq.~(\ref{eq:normalization-condition}). Then the action of $\mathbf{R}$ on Eq.~(\ref{eq:S-on-phi-n}) is given by Eq.~(\ref{eq:reset-action}). The probability of deriving $\supket{m_1^S m_2^S\ldots m_{|S|}^S 11\ldots 1}$ is
\begin{align}
    \sum_{m^I}p_{m^S,m^I}= \sum_{m^I}\supbra{\phi_{m^S,m^I}^{ (0)}}S_{\lambda,\Delta t} \supket{\phi_n^{ (0)}}.
\end{align}
Note that the summation  $\sum_{m^I}$ includes terms from  (\ref{eq:diagonal-matrix-element}) and the second line of  (\ref{eq:off-diagonal-matrix-element}) with all $i\in I$. Thus in the expression of the above probability, terms concerning $B_{n,i}$ cancel; in other words, they have no contribution to the whole evolution. So the property is proved.\hfill $\blacksquare$\par

The evolution matrix $S_{\lambda,\Delta t}$ resembles the classical Markov matrix describing infection spreading in networks, such as the one shown in Ref.~\mycite{6}{Mieghem_09}. Specifically, assuming an infinitesimal generator $Q$ of the Markov chain as defined in Ref.~\mycite{6}{Mieghem_09}, the evolution matrix thus correspond to 
\begin{align}
    S_{\lambda,\Delta t}\sim e^{Q^T \Delta t}
\end{align}
where the superscript $T$ denotes matrix transpose.

Repeated actions of the stochastic matrix lead to exponential decay behavior, as we have seen in the numerical simulations in the main text. In the next subsection, we will explicitly calculate the decay rate of a susceptible node when it is coupled to an infectious node.

\subsection{EXAMPLE: One infectious node and one susceptible node}
As an example of the above derivation, we calculate the matrix elements of $ S_{\lambda,\Delta t}$ in a simple network with one infectious node and one susceptible node. The Hamiltonian of this simple network reads
\begin{align}
    H =& -\gamma Z^s_0 Z^s_1\nonumber \\
        & -\lambda   (X^s_0 X^b_0+X^s_1 X^b_1)\nonumber \\
        & -\alpha Z^b_0 Z^b_1,
        \label{eq:simplified-2-2-hamiltonian}
\end{align}
where the infectious node is labelled by $0$ and the susceptible node by $1$. As explained previously, the initial state is a classical bit string. Specifically, the 1 infectious and 1 susceptible system is initialized as $\ket{init}=\ket{0}_1\otimes\ket{1}_0=\ket{01}$. The survival probability of the susceptible node $\hat{P}_1$ is estimated by measuring the observable $Z_1 = Z_1\otimes I_0$(See Eq. (12) in the main text), which can be written as
\begin{align}
    Z_1\otimes I_0=\ket{00}\bra{00}+\ket{01}\bra{01}-\ket{10}\bra{10}-\ket{11}\bra{11},
\end{align}
which are all projectors of the classical bit string. For simplicity, we represent classical configurations of bit string utilizing column vectors
\begin{align}
    \supket{\phi_{00}^{ (0)}}=\left ( \begin{array}{c}
        1 \\
        0 \\
        0 \\
        0 \end{array} \right),
        \supket{\phi_{01}^{ (0)}}=\left ( \begin{array}{c}
        0 \\
        1 \\
        0 \\
        0 \end{array} \right),\nonumber\\
        \supket{\phi_{10}^{ (0)}}=\left ( \begin{array}{c}
        0 \\
        0 \\
        1 \\
        0 \end{array} \right),
        \supket{\phi_{11}^{ (0)}}=\left ( \begin{array}{c}
        0 \\
        0 \\
        0 \\
        1 \end{array} \right).
\end{align}
A classical probability distribution $\{p_{m^S,m^I}\}$ can be written into a column vector using this basis. The matrix representation of $\mathbf{R}$ is
\begin{align}
    \mathbf{R}=\left ( \begin{array}{cccc}
        0 & 0 & 0 & 0 \\
        1 & 1 & 0 & 0  \\
        0 & 0 & 0 & 0  \\
        0 & 0 & 1 & 1  \end{array} \right)
\end{align}
so that
\begin{align}
    \left ( \begin{array}{c}
        p_{00} \\
        p_{01} \\
        p_{10} \\
        p_{11} \end{array} \right)\stackrel{\mathbf{R}}{\longrightarrow}
        \left ( \begin{array}{c}
        0 \\
        p_{01}+p_{00} \\
        0 \\
        p_{11}+p_{10} \end{array} \right)=\mathbf{R}\left ( \begin{array}{c}
        p_{00} \\
        p_{01} \\
        p_{10} \\
        p_{11} \end{array} \right).
\end{align}
According to Eq.~(\ref{eq:spectrum-function}). The spectrum functions are given by
\begin{align}
    \tilde{h}^{00} (\omega)&=\tilde{h}^{11} (\omega)=2\pi\delta (\omega+2\alpha),\nonumber\\
    \tilde{h}^{01} (\omega)&=\tilde{h}^{10} (\omega)=0.
\end{align}
The matrix elements of $S_{\lambda,\Delta t}$ read
\begin{align}
    S_{\lambda,\Delta t}=&\left ( \begin{array}{cccc}
        1-2\lambda^2 A_0 & \lambda^2 A_1 & \lambda^2 A_1 & 0 \\
        \lambda^2 A_0 & 1-2\lambda^2 A_1 & 0 & \lambda^2 A_0  \\
        \lambda^2 A_0 & 0 & 1-2\lambda^2 A_1 & \lambda^2 A_0  \\
        0 & \lambda^2 A_1 & \lambda^2 A_1 & 1-2\lambda^2 A_0  \end{array} \right)\nonumber\\
        &+\mathcal{O} (\lambda^4),
\end{align}
where
\begin{align}
    A_0\equiv \frac{\sin^2 ( (\gamma+\alpha)\Delta t)}{ (\gamma+\alpha)^2},\quad A_1\equiv \frac{\sin^2 ( (\gamma-\alpha)\Delta t)}{ (\gamma-\alpha)^2}.
\end{align}
It can be seen explicitly that both  $S_{\Delta t}$ and $\mathbf{R}$ are stochastic matrices, satisfying the relationship
\begin{align}
    \sum_m  (S_{\lambda,\Delta t})_{mn}=1,\quad  (S_{\lambda,\Delta t})_{mn}\geq 0,
\end{align}
for small enough $\lambda$. The effect of staggering implementations $\mathbf{R}S_{\lambda,\Delta t}\mathbf{R}S_{\lambda,\Delta t}\ldots$ on an initial state can be evaluated by assuming the following stochastic matrix's eigendecomposition
\begin{align}
    \mathbf{R}S_{\lambda,\Delta t}=U\left ( \begin{array}{cccc}
        1 & 0 & 0 & 0 \\
        0 & 1-\lambda^2 (A_0+A_1) & 0 & 0  \\
        0 & 0 & 0 & 0  \\
        0 & 0 & 0 & 0  \end{array} \right)U^{-1},
        \label{eq:eigen-decomposition}
\end{align}
where the columns of $U$ are the corresponding eigenvectors of $ \mathbf{R}S_{\lambda,\Delta t}$. The only non-zero eigenvalue $1-\lambda^2 (A_0+A_1)$ leads to the single-exponential decay behaviour in the numerical simulation. The expectation value of the bit string projector $\ket{l}\bra{l}$ given an initial state $\ket{k}\bra{k}$ after evolving time $t$ is 
\begin{equation}
\begin{aligned}
     &\supbra{\phi_{l}^{ (0)}} (\mathbf{R}S_{\lambda,\Delta t})^{t/\Delta t} \supket{\phi_{k}^{ (0)}}\\
     =&B_0+B_1[1-\lambda^2 (A_0+A_1)]^{t/\Delta t},
\end{aligned}
\end{equation}
where $B_0,B_1$ are some irrelevant $t$-independent constants. To match the exponential decay behaviour shown in the main text, the above formula can be reformulated as $B_0+B_1e^{-\Gamma t}$. The decay constant on the exponential is the infection rate, which reads 
\begin{equation}
\begin{aligned}
    \Gamma (\gamma)=&\frac{1}{\Delta t}\ln\frac{1}{1-\lambda^2 (A_0+A_1)}\\
    =&\frac{\lambda^2 (A_0+A_1)}{\Delta t}+\mathcal{O} (\lambda^4).
    \label{Gamma}
\end{aligned}
\end{equation}
Compare the coefficients $A_0$ and $A_1$. Note that $A_0$ and $A_1$ are like a Dirac delta function according to the identity
\begin{align}
    \delta (\omega)=\lim_{\Delta t\rightarrow\infty}\frac{\sin^2 (\omega\Delta t)}{\pi \omega^2\Delta t}.
\end{align}
Thus for $\alpha\gg |\alpha-\gamma|$ and large $\Delta t$, $A_0$ can be neglected compared with $A_1$. It leaves
\begin{align}
    \Gamma (\gamma)\simeq \lambda^2\Delta t \left (\frac{\sin ( (\gamma-\alpha)\Delta t)}{ (\gamma-\alpha)\Delta t}\right)^2.
    \label{eq:infection-rate-analytic}
\end{align}
Thus derived sinc-function has been checked in the numerical simulation of Fig. 2 and 3 in the main text. 

Eq.~(\ref{eq:infection-rate-analytic}) shows that the infection rate has the following two properties. (1) If the system and bath Hamiltonian are identical, i.e., $\gamma=\alpha$, $\Gamma$ will reach its maximum $\lambda^2\Delta t$. (2) If we set $\alpha\Delta t=k\pi,k=1,2,\ldots$,the infection rate $\Gamma\rightarrow 0$ as $\gamma\rightarrow 0$. The first property shows that there exists a maximum infection rate by tuning the inter-system coupling $\gamma$. We can use this inter-system coupling to simulate the household transmission in the numerical simulation. The second property indicates that the long-range transmission process, where the infection rate is small, can be simulated using the thermal dynamic model.

The results of network with one infectious node and one susceptible node can be generalized to the network connecting multi-infectious nodes and one susceptible node. According to the independence of the infectious nodes and the expression for $A_{n,j}$ in Eq.~(\ref{eq:Anj-Bnj}), the infection rate of the susceptible node $j$ is 
\begin{align}
    \Gamma_j\simeq \lambda^2\Delta t\left (\frac{\sin (\sum_{i'\in I} (\gamma_{i'j}-\alpha_{i'j})\Delta t)}{\sum_{i'\in I} (\gamma_{i'j}-\alpha_{i'j})\Delta t}\right)^2.
    \label{eq:multi-infectious-Gamma}
\end{align}
This formula shows explicitly that if we take $\sum_{i'\in I}\alpha_{i'j}\Delta t=k\pi$ and $\gamma_{i'j}\rightarrow 0$, the infection rate tends to zero up to $\mathcal{O} (\lambda^2)$, as proved in the previous subsection. The phenomenological SI model requires the linearity of the infection rate $\Gamma_j (\sum_{i'\in I}\gamma_{i'j})=\sum_{i'\in I}\Gamma_j (\gamma_{i'j})$. However, due to the non-linearity of the sinc-function, Eq.~(\ref{eq:multi-infectious-Gamma}) can only satisfy
\begin{equation}
\begin{aligned}
     \Gamma_j (\sum_{i'\in I}\gamma_{i'j})&=\frac{\lambda^2\Delta t}{k^2}\left (\sum_{i'\in I}\frac{\sin (\gamma_{i'j}\Delta t)}{\pi}\right)^2+\mathcal{O} (\gamma^3)\\
     &=\left (\sum_{i'\in I}\frac{1}{k}\sqrt{\Gamma (\gamma_{i'j})}\right)^2+\mathcal{O} (\gamma^3)
     \label{eq:non-linearity}
\end{aligned}
\end{equation}
where we take $\alpha\Delta t=\pi$ in $\Gamma (\gamma_{i'j})$ and $\sum_{i'\in I}\alpha_{i'j}\Delta t=k\pi$ in $\Gamma_j$. Thus, the infection rate of $j$ surrounded by multi-infectious nodes qualitatively satisfies the requirement of phenomenology if we choose $k=1$ in the numerical simulation. Taking $k=1$ and requiring a uniform inter-bath coupling, we have
\begin{align}
    \alpha_{i'j}=\alpha=\frac{\pi}{|I|\Delta t}.
    \label{eq:multi-infectious-bath-coupling}
\end{align}
This setting is utilized in our numerical simulation with two index patients, as shown in Fig. 7 in the main text.

\section{Numerical results using explicit fourth-order Runge-Kutta method}\label{app:Numerical-results-using-exact-diagonalization-method}

Numerical results in the main text are from the quantum simulation, which have errors by first-order Trotter decomposition and finite number of measurements. To check if the error in quantum simulation is small enough, we use the Runge-Kutta method to carry out the Sch\"odinger evolution given an initial density operator $\rho (0)$. The Runge-Kutta method is a family of methods of solving partial differential equations numerically. We utilize the fourth-order Runge-Kutta method to solve Schr\"odinger evolution on density operator
\begin{equation}
    i\frac{\partial \rho (t)}{\partial t}=[H,\rho (t)].
\end{equation}
$N$-th order Runge-Kutta method accumulates error in the order of $O (h^{N+1})$ in each step, where $h$ is the step time adopted~\mycite{39}{DORMAND198019}. In our numerical calculation, step time $h$ is as small as the machine precision thus the systematic error in the numerical results can be neglected. However, this method can only simulate small networks. Since the number of partial differential equations equals the number of entries in the density matrix, the time complexity of the method grows exponentially as the number of qubits increases.

\begin{figure}
    \centering
  \includegraphics[width = 0.45\textwidth]{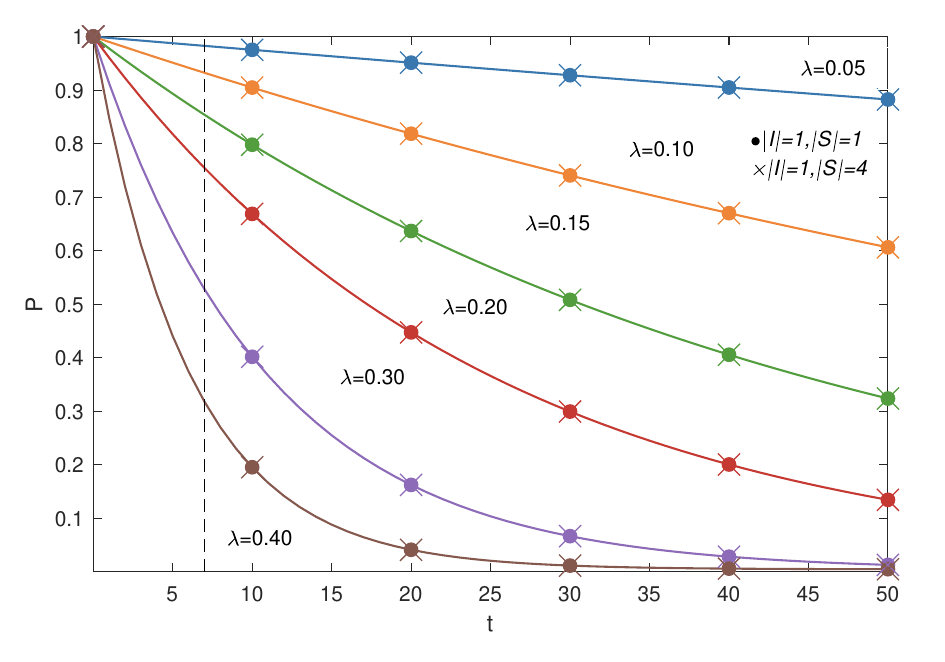}
  \includegraphics[width = 0.45\textwidth]{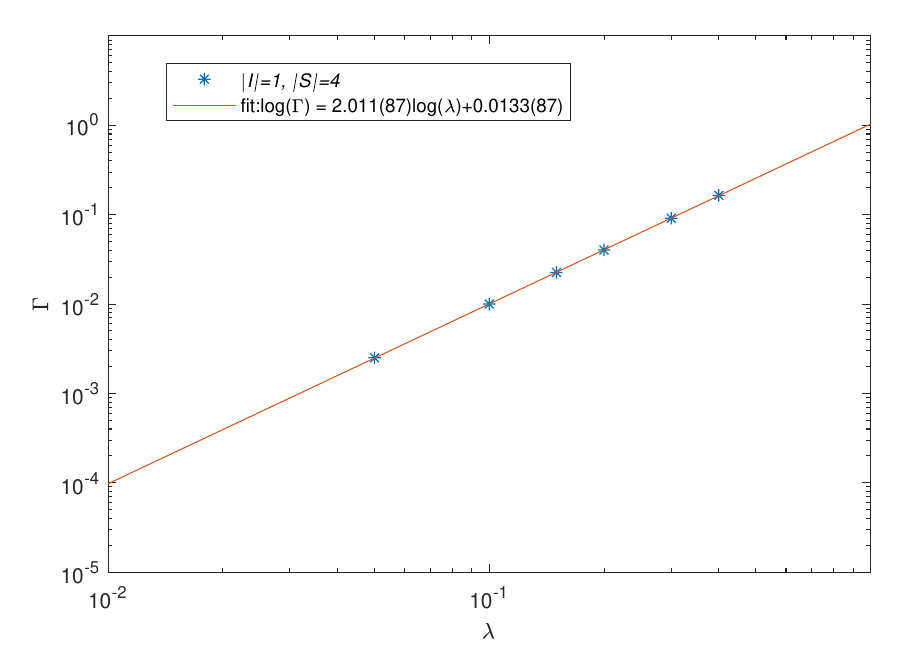}
  \caption{The survival probability of the household node 1 as a function of time is shown in the left panel, and the infection rate $\Gamma$ of the household node as a function of system-bath coupling $\lambda$ is shown in the right panel. These numerical results are derived from the fourth-order Runge-Kutta method that can be cross-checked with those in Fig. 2 in the main text. We find the same single exponential behaviour of survival probabilities on the household node and the same $\Gamma$ dependence on $\gamma$.}
  \label{fig:Runge-Kutta_infection-rate-system-bath-coupling}
\end{figure}

In Fig.~\ref{fig:Runge-Kutta_infection-rate-system-bath-coupling} and Fig. 6 in the main text, we use fourth-order Runge-Kutta method to check with results by the first-order Trotter decomposition in the main text Fig. 2 and Fig. 6 respectively. We find that results from the quantum simulation converge to those of the Runge-Kutta method within the error of statistics. It indicates that errors in the quantum simulation are well controlled. 

\begin{center}
   \line(1,0){510}
\end{center}
 
\bibliographystyle{unsrt}
\nobibliography{main.bib}
